\documentclass[11pt, a4paper]{article}

\usepackage[utf8]{inputenc}
\usepackage{amsmath}
\usepackage{amsfonts}
\usepackage{amssymb}
\usepackage{graphicx}
\usepackage{listings}
\usepackage{xcolor}
\usepackage{hyperref}
\usepackage[margin=1in]{geometry}
\usepackage{graphicx}
\usepackage{textgreek}
\usepackage{subcaption} 
\usepackage[font=small,labelfont=bf]{caption}
\usepackage[superscript]{cite}
\usepackage{authblk}
\usepackage{lineno}

\definecolor{codebg}{RGB}{245, 248, 252}        
\definecolor{codecomment}{RGB}{80, 115, 150}    
\definecolor{codekeyword}{RGB}{0, 85, 170}      
\definecolor{codestring}{RGB}{0, 128, 160}     
\definecolor{codenumber}{RGB}{140, 160, 180}    
\definecolor{codeidentifier}{RGB}{15, 35, 75}   

\usepackage{authblk}
\title{\textbf{A Computational and Experimental Framework for Fourier Optics in Python}}
\author[1]{Rachel Migdal} 
\author[1]{Antoine Islegen-Wojdyla} 
\affil[1]{Advanced Light Source, Lawrence Berkeley National Laboratory}

\date{November 18, 2026}

\begin{document}

\maketitle

\begin{abstract}

Simulation of the propagation of light is useful for the design and dimensioning of optical systems. In this paper, we present the basics of propagation simulation using Fourier optics and practical implementation in Python, with a particular emphasis on dimensions. We first verify that our numerical wave-optics simulations maintain correct physical spatial scaling. Next, we benchmark numerical propagation against closed-form analytical formulas for Gaussian beams, lens focusing effects, and grating diffraction. We then show  how those simulations can be used to generate diffractive optical elements such as Fresnel zone plates, and we bridge the gap between numerical simulations and experiments by collecting data and recovering phase information using a phase retrieval algorithm.
\end{abstract}


\section{Introduction} 
Light propagates through space and time as an electromagnetic wave, whose behavior can in principle be described rigorously via Green's function formalisms. In many scenarios relevant to optical imaging and laser systems, however, a complete vector wave treatment is unnecessary. Instead, the propagation is well described by Fourier optics—a frequency-based approximation that emerges naturally when propagation distances are large and the system remains relatively close to the optic axis (the paraxial regime).

\begin{figure}[htbp]
    \centering
    \begin{minipage}{0.9\textwidth}
        \centering
        \includegraphics[width=\textwidth]{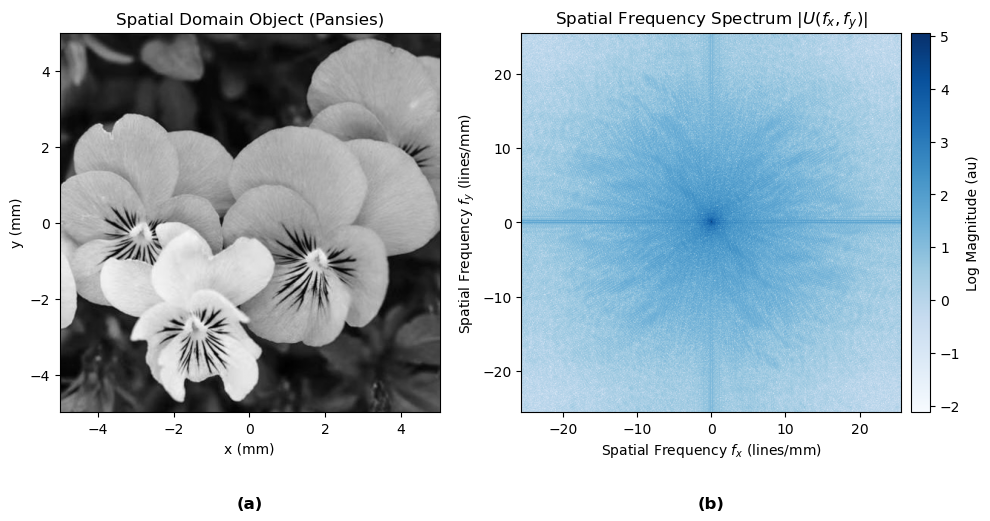}
        \caption{Conceptual representation of spatial frequency decomposition in Fourier optics: \textbf{(a)} An image of pansies in physical spatial coordinates; \textbf{(b)} The corresponding 2-D spatial frequency magnitude spectrum (displayed on a logarithmic scale), where each frequency point physically represents a plane wave traveling at a unique propagation angle.}
        \label{fig:spatial_freq}
    \end{minipage}
\end{figure}

At its core, this framework relies on the property of linearity, allowing us to regard optical propagation as a mapping between an object's spatial distribution $g(x,y)$ and its corresponding spatial frequency spectrum $G(f_x, f_y)$, defined via the two-dimensional Fourier transform:
\begin{equation}
    G(f_x, f_y) = \int_{-\infty}^{\infty} \int_{-\infty}^{\infty} g(x,y) e^{-j2\pi(f_x x + f_y y)} \,dx\,dy
\end{equation}
In this representation, each frequency point physically corresponds to a plane wave traveling at a unique propagation angle, as illustrated in Figure \ref{fig:spatial_freq}. Where applicable, this Fourier approach drastically simplifies complex geometries and optical elements into manageable mathematical expressions that can be computed efficiently. Beyond computational convenience, it provides profound physical insight into the behavior of imaging systems, diffraction limits, and spatial frequency filtering.

To bridge the gap between theoretical wave propagation and practical simulation, this paper presents the fundamentals of Fourier optics simulation implemented in Python, with a rigorous emphasis on spatial scaling. We begin by verifying that our numerical wave-optics framework maintains correct physical scaling, and we benchmark our propagation routines against closed-form analytical formulas for Gaussian beams, lens focusing effects, and grating diffraction. Building on these validated foundations, we demonstrate how these simulations can be used to generate diffractive optical elements such as Fresnel zone plates. Finally, we connect numerical models directly to experimental work by processing real data and recovering lost phase information via phase retrieval algorithms.

A working familiarity with this frequency-space description of real-space objects will serve as our primary mathematical language as we move forward. Keeping these tools in mind will take us a long way through the derivations and numerical implementations that follow in Section 2.

\section{Theoretical Background: Fourier Analysis in Optics}
\subsection{The Fourier Transform as a Mathematical Decomposition}
The primary utility of the Fourier transform (FT) in optics is its ability to perform a mathematical decomposition. Building upon the general two-dimensional transform introduced in Section 1, we specialize this relationship for optical fields. A complex scalar field distribution $g(x,y)$ evaluated at an input plane can be broken down into a continuous linear combination of elementary plane waves via the forward and inverse Fourier transform pair:

\begin{align}
    G(f_x, f_y) &= \int_{-\infty}^{\infty} \int_{-\infty}^{\infty} g(x,y) \, e^{-j2\pi(f_x x + f_y y)} \,dx\,dy \\
    g(x,y) &= \int_{-\infty}^{\infty} \int_{-\infty}^{\infty} G(f_x, f_y) \, e^{j2\pi(f_x x + f_y y)} \,df_x\,df_y
\end{align}

where $G(f_x, f_y)$ determines the relative complex amplitude of each specific frequency component. 

\begin{figure}[htbp]
    \centering
    \begin{minipage}{0.6\textwidth}
        \centering
        \includegraphics[width=\textwidth]{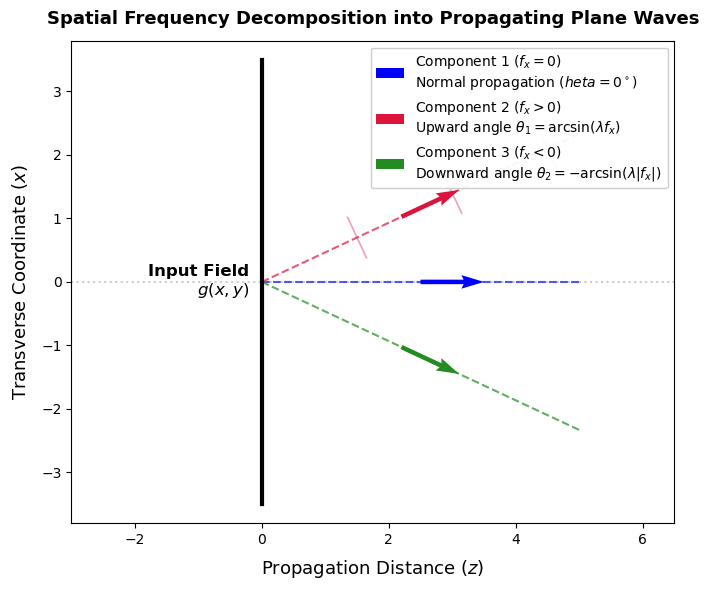}
        \caption{Conceptual schematic of spatial frequency decomposition in Fourier optics. An arbitrary input field distribution $g(x,y)$ at $z = 0$ is decomposed into a linear superposition of plane waves traveling at distinct propagation angles $\theta = \arcsin(\lambda f_x)$ determined by their spatial frequency components $f_x$.}
        \label{fig:planewave}
    \end{minipage}
\end{figure}

Physically, each term in the inverse integral can be identified as an elementary function consisting of a plane wave traveling in a unique direction away from the input plane, as illustrated in Figure \ref{fig:planewave}. The spatial frequencies $(f_x, f_y)$ define the transverse wavelength and propagation angles via the directional cosines:
\begin{equation}
    f_x = \frac{\sin\theta_x}{\lambda}, \quad f_y = \frac{\sin\theta_y}{\lambda}
\end{equation}

This decomposition allows us to calculate the response of an optical system to an arbitrary complex field by simply superimposing the independent contributions of these individual plane wave eigenmodes, as demonstrated schematically by the continuous plane-wave fanout in Figure \ref{fig:planewave}.

\begin{figure}[htbp]
    \centering
    \begin{minipage}{1\textwidth}
        \centering
        \includegraphics[width=\textwidth]{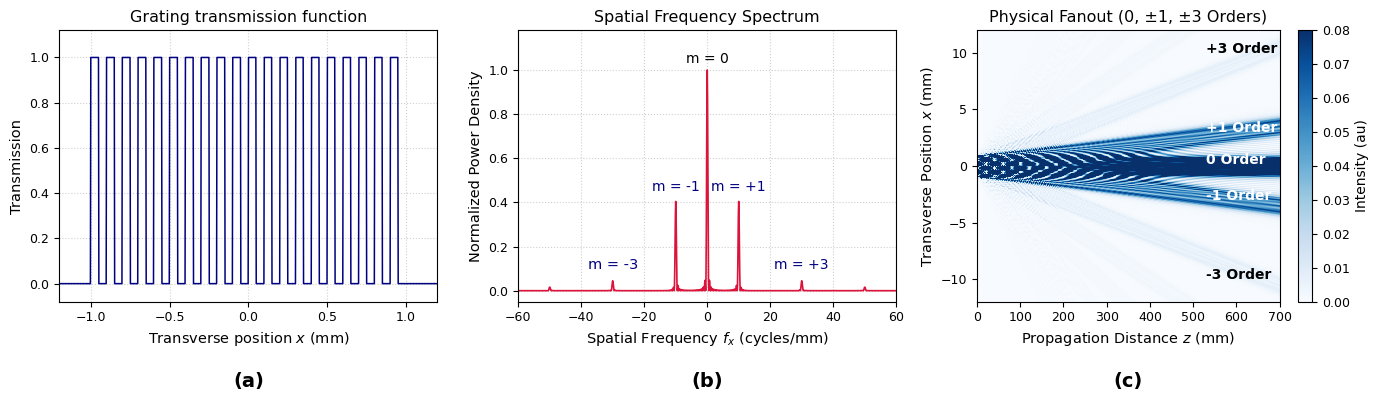}
        \caption{Physical interpretation of Fourier decomposition via a periodic diffraction grating: \textbf{(a)} A plane-wave optical beam at normal incidence interacts with a grating of period $d$, acting as a spatial phase/amplitude filter; \textbf{(b)} The periodic structure decomposes the incident wavefront into a discrete superposition of plane wave eigenmodes, which physically fan out into discrete diffracted orders ($0, \pm 1, \dots$); \textbf{(c)} Propagation fanout into distinct, angularly separated diffracted orders along the $z$-axis.}
        \label{fig:gratingsuperpos}
    \end{minipage}
\end{figure}

To see how this continuous decomposition applies to periodic structures, consider what happens when an optical field interacts with a repeating grating. Rather than a continuous spectrum, periodic boundaries constrain the decomposition to a discrete set of spatial frequencies. This behavior is illustrated in Figure \ref{fig:gratingsuperpos}, where an incident beam interacting with a periodic transmission function is decomposed into discrete plane wave eigenmodes that fan out into distinct, angularly separated diffraction orders ($0, \pm 1, \dots$).

\section{Computational Foundations: Numerical Grid and Frequency Analysis} 
The foundation of a wave-optics simulation is the translation of continuous physical coordinates into discrete numerical arrays defined in terms of physical quantities, such as meters. This environment is established using a uniformly spaced spatial coordinate array $x$ spanning $[-L/2, L/2]$, which follows the standard index arrangement where a finite side length $L$ is spanned by $M$ samples. The relationship between these parameters is defined by:
\begin{equation}
L = M \Delta x
\end{equation}
where $\Delta x$ is the spatial sampling interval. In two dimensions, this is extended to a grid using \texttt{np.meshgrid}:
\begin{equation}
x = \text{linspace}\left(-\frac{L}{2}, \frac{L}{2}, \frac{L}{\Delta x} + 1\right)
\end{equation}

\subsection{Transition to the Frequency Domain via the DFT}
To move from continuous wave theory to numerical simulation, we utilize the Discrete Fourier Transform (DFT), which approximates the analytic Fourier integral through a Riemann sum. This Fourier framework is essential because it allows us to treat optical phenomena as linear spatial filters, decomposing complex field distributions into simpler "elementary" plane waves. 

By convention, the frequency domain is divided into $M$ and $N$ evenly spaced coordinate values. This is accomplished by the following substitutions for spatial frequencies $f_x$ and $f_y$:

\begin{equation}
\begin{gathered}
f_x \rightarrow \frac{p}{M \Delta x}, \qquad \text{and} \qquad f_y \rightarrow \frac{q}{N \Delta y} \\
\text{where} \qquad p = - \frac{M}{2}, \cdots , \frac{M}{2} - 1, \qquad \text{and} \qquad q = - \frac{N}{2}, \cdots , \frac{N}{2} - 1 \\
    \Delta f_x = \frac{1}{M \Delta x}=\frac{1}{L_x}, \qquad \text{and} \qquad \Delta f_y = \frac{1}{N \Delta y}=\frac{1}{L_y}
\end{gathered}
\end{equation}

The parameters $\Delta f_x$ and $\Delta f_y$ represent the "frequency sample intervals," which dictate the spectral resolution of the simulation.

\subsection{Fast Fourier Transform (FFT) and Array Centering}
Because optical simulations frequently require cascading multiple propagation steps or performing repeated field transformations, we utilize the Fast Fourier Transform (FFT) algorithm \cite{goodman1996,voelz2011}. The FFT is a highly efficient implementation of the Discrete Fourier Transform (DFT) that reduces the computational complexity from $O(N^2)$ to $O(N \log_2 N)$ complex multiplications and additions \cite{voelz2011}. 

To appreciate this computational gain, consider a 1-D spatial coordinate array of $N = 1000$ grid points. A direct matrix evaluation of the DFT requires $N^2 = 1,000,000$ operations, whereas the FFT algorithm requires only $N \log_2 N \approx 1000 \times 10 = 10,000$ operations—achieving a $100\times$ speedup \cite{voelz2011}. This advantage scales even more dramatically in two-dimensional simulations ($N \times N$ grids). For a modest $1000 \times 1000$ spatial array ($10^6$ total pixels), a naive 2-D DFT demands $O(N^4)$ or $(10^6)^2 = 10^{12}$ (one trillion) operations, rendering standard desktop evaluation completely unrealistic. In contrast, the 2-D FFT executes in $O(N^2 \log_2 N)$ or roughly $2 \times 10^7$ operations. This mathematical efficiency makes 2-D wave-optics simulations accessible on standard personal computers rather than requiring high-performance supercomputing clusters \cite{voelz2011}.

Mathematically, the Fourier transform is essentially its own inverse (a near-involution property) \cite{goodman1996}. The forward Fourier transform $\mathcal{F}$ and the inverse Fourier transform $\mathcal{F}^{-1}$ share identical functional structures, differing only by a sign flip in the complex exponential kernel and an overall normalization factor $1/N$ \cite{goodman1996, voelz2011}:
\begin{equation}
    \mathcal{F}\{u(x)\} = \int_{-\infty}^{\infty} u(x) \exp(-\mathrm{j} 2\pi f_x x) dx, \qquad \mathcal{F}^{-1}\{U(f_x)\} = \int_{-\infty}^{\infty} U(f_x) \exp(+\mathrm{j} 2\pi f_x x) df_x
\end{equation}
Because of this fundamental symmetry, both forward spectrum decomposition and inverse field reconstruction utilize the exact same FFT computational engine \cite{voelz2011}.

A practical issue in FFT implementation concerns the spatial indexing of array values. While continuous physical optics models are centered around the optical axis ($x = 0$), standard FFT libraries (such as \texttt{numpy.fft}) expect the zero-frequency coordinate to reside at the first index position (\texttt{}). To maintain physical consistency and prevent phase-wrapping artifacts during numerical propagation, \texttt{np.fft.fftshift} and \texttt{np.fft.ifftshift} are applied to re-center both the complex field arrays and the spatial-frequency coordinate bins ($f_x$) \cite{voelz2011}.

Free-space optical propagation is modeled in the spatial-frequency domain via the angular spectrum / transfer function approach (\texttt{propTF}). Given an input complex electric field $U_{\text{in}}(x)$ defined across a transverse window size $L$, wavelength $\lambda$, and propagation distance $z$, the propagated output field $U_{\text{out}}(x)$ is evaluated through the following three-step procedure:

\begin{enumerate}
    \item \textbf{Complex Angular Spectrum Computation:} Convert the input spatial field $u_{\text{in}}(x)$ into the spatial-frequency domain using the forward Fast Fourier Transform:
    \begin{equation}
        U_{\text{in}}(f_x) = \mathcal{F}\left\{ u_{\text{in}}(x) \right\},
    \end{equation}
    where $f_x$ denotes the spatial frequency coordinate and $\mathcal{F}$ represents the 1-D Fourier transform operator.

    \item \textbf{Transfer Function Modulation:} Apply the paraxial (Fresnel) free-space transfer function $H(f_x)$ to introduce the distance-dependent quadratic phase shift across spatial frequencies \cite{goodman1996, voelz2011}:
    \begin{equation}
        H(f_x) = \exp\left( -\mathrm{j} \pi \lambda z f_x^2 \right),
    \end{equation}
    where $\mathrm{j} = \sqrt{-1}$ is the imaginary unit.

    \item \textbf{Field Reconstruction:} Transform the modulated spectrum back to the spatial domain via the inverse Fast Fourier Transform:
    \begin{equation}
        U_{\text{out}}(x) = \mathcal{F}^{-1}\left\{ U_{\text{in}}(f_x) \cdot H(f_x) \right\}.
    \end{equation}
\end{enumerate}

\subsection{Intensity and the Optical Phase Problem}

While the wave propagation operator \texttt{propTF} processes the full complex electric field $U_{\text{out}}(x) = |U_{\text{out}}(x)| \exp[\mathrm{j}\phi(x)]$, physical sensors cannot respond to the ultra-high oscillation frequencies of optical fields ($\sim 10^{14}\text{ Hz}$) \cite{goodman1996, voelz2011}. Consequently, real-world optical detectors measure only the time-averaged irradiance, or intensity $I(x)$ \cite{goodman1996}:
\begin{equation}
    I(x) = |U_{\text{out}}(x)|^2 = U_{\text{out}}(x) \cdot U_{\text{out}}^*(x)
\end{equation}
where $U_{\text{out}}^*(x)$ denotes the complex conjugate. 

In this measurement process, the spatial phase distribution $\phi(x) = \arg(U_{\text{out}}(x))$ is completely discarded. The loss of phase information constitutes the fundamental phase problem in optics \cite{goodman1996, miyakawa2014}. Because wavefront curvature, focus position, and spatial aberrations are encoded directly within the phase profile, intensity-only measurements make inverse optical problems—such as wavefront sensing and aberration correction—inherently challenging, necessitating specialized recovery techniques such as interferometry, holography, or iterative phase retrieval \cite{goodman1996, miyakawa2014, zalevsky1996}.

\begin{figure}
    \centering
    \includegraphics[width=1\linewidth]{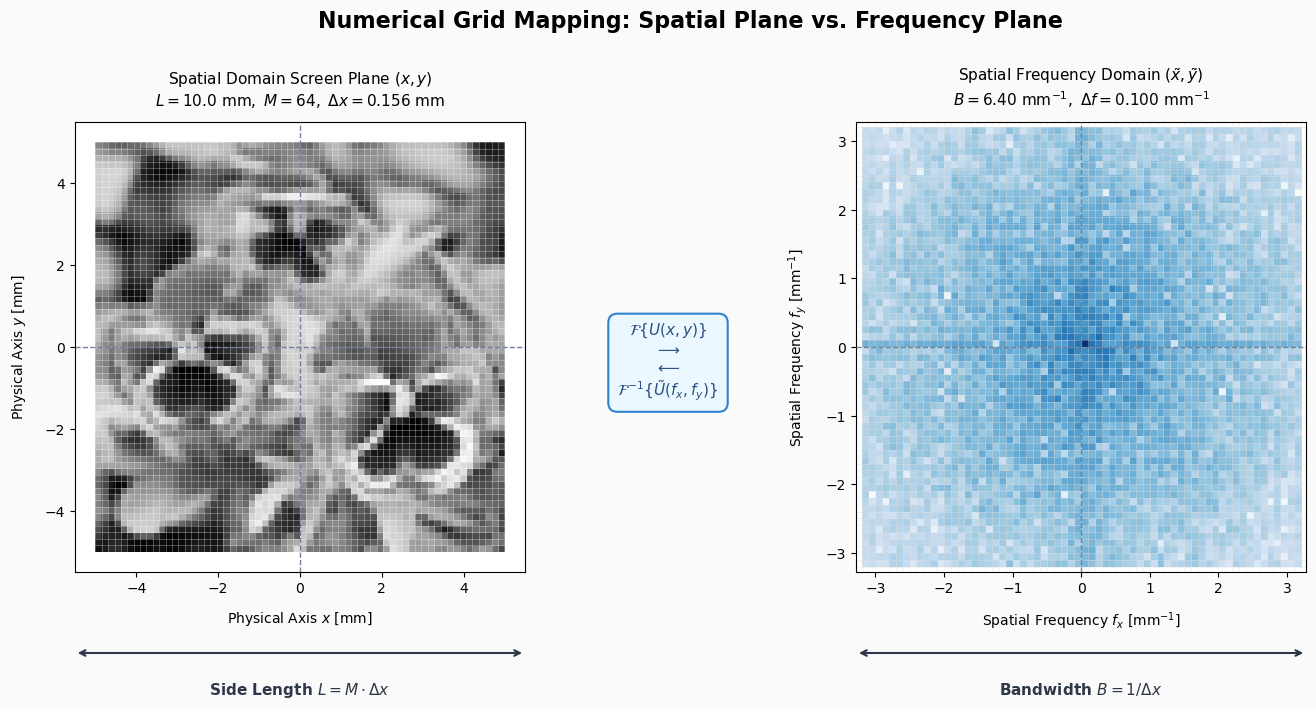}
    \caption{Comparison between the spatial screen plane $(x, y)$ and the spatial frequency plane $(f_x, f_y)$ for an $M \times M$ grid. The spatial domain spans a physical window length $L = M \cdot \Delta x$, where $\Delta x$ represents the spatial sampling interval. Under the 2D Fast Fourier Transform ($\mathcal{F}$), the corresponding frequency spectrum spans a bandwidth $B = 1/\Delta x$ with a frequency sampling resolution of $\Delta f = 1/L$.}
    \label{fig:placeholder}
\end{figure}

The implementation accepts the complex input field distribution $U_{\text{in}}(x)$, optical wavelength $\lambda$, numerical window size $L$, and displacement $z$, returning the complex optical field $U_{\text{out}}(x)$ at the observation plane.

\begin{lstlisting}[caption={Defining the Foundational Fourier Transform Function}]
def propTF(uin_V_m, L_m, lambda_m, z_m):
    M = uin_V_m.size        # number of samples (grid points)
    dx_m = L_m/M            # grid spacing
    k_1_m = 2*np.pi/lambda_m # wavenumber

    # Frequency coordinates
    fx_1_m = np.linspace(-1/(2*dx_m), 1/(2*dx_m) - (1/L_m), M)
    
    # Transfer function for Fresnel propagation
    H = np.exp(-1j * np.pi * lambda_m * z_m * (fx_1_m**2))
    H = np.fft.fftshift(H)

    # Fourier transform input field
    Uin_V_m = np.fft.fft(np.fft.fftshift(uin_V_m))

    # Multiply in Fourier space
    Uout_V_m = H * Uin_V_m

    # Inverse FFT --> propagated field
    uout_V_m = np.fft.ifftshift(np.fft.ifft(Uout_V_m))
    return uout_V_m
\end{lstlisting}

\subsection{Elementary Optical Fields}

The transverse mode is simulated as a Gaussian intensity distribution:
\begin{equation}
g(x) = \exp\left(-\frac{x^2}{2\sigma^2}\right)
\end{equation}
where $\sigma$ represents the beam Root Mean Square (RMS) width. We note that the RMS spatial parameter is distinct from the standard deviation, as the latter pertains strictly to statistical probability distributions rather than physical spatial intensity distributions.

In optics, it is also common practice to characterize the beam width using the Full Width at Half Maximum ($\text{FWHM}$). For an ideal Gaussian intensity distribution, there exists a constant linear relationship between the RMS width $\sigma$ and the $\text{FWHM}$:
\begin{equation}
\text{FWHM} = 2\sqrt{2\ln 2}\,\sigma \approx 2.355\sigma
\end{equation}
While both quantities can be experimentally measured, the $\text{FWHM}$ is often simpler to measure directly from intensity profiles using standard beam profilers, whereas the RMS width provides a direct link to energy distribution and second-moment beam parameters.

\begin{lstlisting}[caption={Defining a 1D Gaussian Beam Intensity Profile}]
    dx_m = 1e-6                                           # grid size, m:meter
    Dx_m = 1                                              # screen size
    x_m = np.linspace(-Dx_m/2, Dx_m/2, int(Dx_m/dx_m)+1) # spatial grid along x-axis 
    
    x_mean_m = 0.0                                       # centered on origin
    x_std_m = 50e-6                                       # RMS beam width (sigma)
    g = np.exp(-(x_m - x_mean_m)**2 / (2 * (x_std_m**2)))  # Gaussian intensity distribution
    
    # Calculate FWHM from RMS
    fwhm_m = 2.355 * x_std_m
\end{lstlisting}

\begin{figure}[htbp]
    \centering
    \begin{subfigure}[b]{0.495\textwidth}
        \centering
        \includegraphics[width=\textwidth]{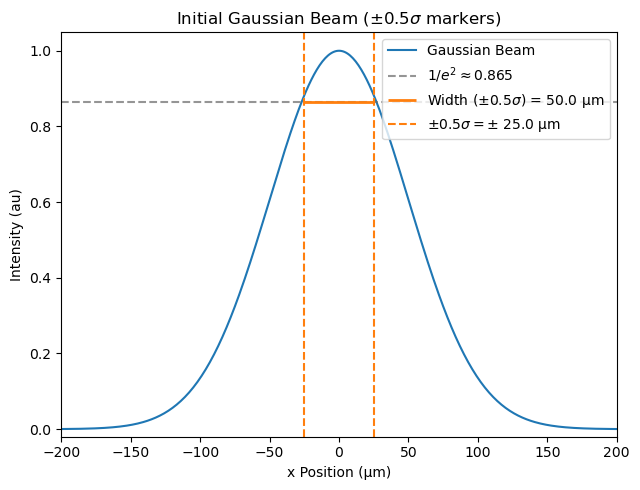}
        \caption{Intensity plot of initial Gaussian beam.}
        \label{fig:initgauss}
    \end{subfigure}
    \hfill
    \begin{subfigure}[b]{0.495\textwidth}
        \centering
        \includegraphics[width=\textwidth]{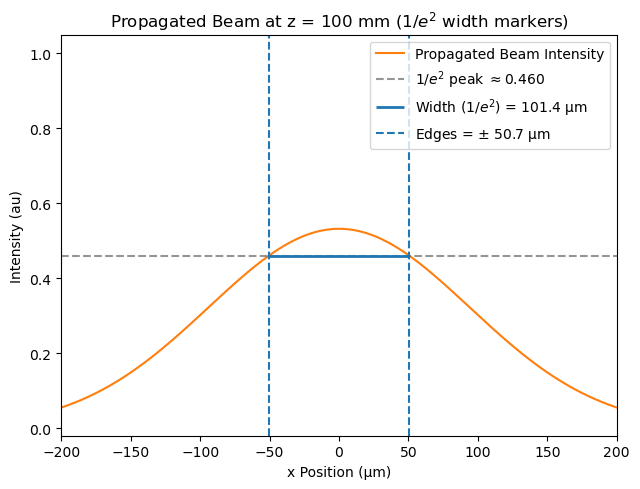}
        \caption{Intensity plot after 100 mm propagation.}
        \label{fig:propgauss}
    \end{subfigure}
    
    \caption{Gaussian beam comparison before and after propagation.}
    \label{fig:both_images}
\end{figure}

The complex optical field $U_0$ is defined as the square root of this intensity to maintain physical consistency during complex propagation. We next propagate the field by $0.1\text{ m}$, compute the resulting intensity, and plot the initial and propagated profiles (Figure \ref{fig:both_images}).

\begin{lstlisting}[caption={Propagating the Gaussian Optical Field}]
    wavelength_m = 500e-9 
    zp_m = 100e-3                             # propagation distance, m
    
    E0 = np.sqrt(g)                           # initial field amplitude (E0^2 = g)
    Ep = propTF(E0, Dx_m, wavelength_m, zp_m) # propTF propagates field by distance zp_m 
    
    Ip = np.abs(Ep)**2                        # propagated intensity 
\end{lstlisting}


As predicted by classical optics principles, the beam diverges when propagated over a distance—demonstrated by the decreased peak intensity and increased width of the Gaussian in Figure \ref{fig:propgauss}.

For a pure Gaussian beam, conservation of the optical étendue determines the angular spread of the beam. The RMS divergence angle $\theta$ can be expressed directly in terms of the wavelength $\lambda$ and initial RMS waist $\sigma_0$:
\begin{equation}
\theta = \frac{\lambda}{4\pi\sigma_0}
\end{equation}
Using this RMS divergence, a good approximation of the beam size at any axial distance $z$ is given by the quadratic sum of the waist size and the spread due to angular divergence\cite{voelz2011}:
\begin{equation}
\sigma(z) = \sqrt{\sigma_0^2 + (\theta z)^2}
\end{equation}

We can generate the same 1-D Gaussian field used earlier and propagate it over a symmetric range of distances ($-1.1\text{ m}$ to $+1.1\text{ m}$) using the Fresnel transfer function \texttt{propTF}. We then extract the RMS beam size from the simulated intensity ($\sigma_{\text{sim}}(z)$) and compare it against the theoretical value.

\begin{lstlisting}[caption={Symmetric Gaussian Beam Propagation}]      
    x_m = np.linspace(-Dx_m/2, Dx_m/2, int(Dx_m/dx_m) + 1)
    
    sigma0_m = 50e-6                               # 50 um RMS waist
    I0 = np.exp(-(x_m**2) / (2 * sigma0_m**2))     # intensity profile
    E0 = np.sqrt(I0)                               # field amplitude (real valued)
    
    Z_m = np.linspace(-1.1, 1.1, 100)   
    
    sim_beam_sizes_m = []
    Ixz = np.zeros((len(Z_m), len(x_m)))           # normalized intensity vs (z,x)
    
    for idx, z in enumerate(Z_m):
        Ez = propTF(E0, Dx_m, wavelength_m, z)
        Iz = np.abs(Ez)**2
    
        # Normalise each slice for visualisation (max = 1)
        Iz_norm = Iz / Iz.max()
        Ixz[idx, :] = Iz_norm
    
        # RMS beam size from intensity weighted second moment
        x_mean = np.sum(x_m * Iz) / np.sum(Iz)
        sigma_z = np.sqrt(np.sum((x_m - x_mean)**2 * Iz) / np.sum(Iz))
        sim_beam_sizes_m.append(sigma_z)
    
    sim_beam_sizes_m = np.array(sim_beam_sizes_m)  
    
    # Analytical RMS beam size 
    theta_rad = wavelength_m / (4 * np.pi * sigma0_m)    # RMS divergence (rad)
    sigma_th_m = np.sqrt(sigma0_m**2 + (theta_rad * Z_m)**2)
\end{lstlisting}
~
~
To determine the numerical divergence angle $\theta$ from our wave propagation simulation, we utilize the asymptotic behavior of the beam envelope. Far from the beam waist (where the propagation distance $z$ is large compared to the initial waist size), the hyperbolic expansion $\sigma(z) = \sqrt{\sigma_0^2 + (\theta z)^2}$ approaches a linear asymptote: $\sigma(z) \approx \theta z$.

Rather than taking a single-point ratio $\sigma/z$, which leaves a residual error due to the initial waist offset $\sigma_0$, we perform a least-squares linear regression (using a first-degree polynomial fit) on the simulated beam sizes across the far-field region ($\vert{}z\vert{} > 0.5\text{ m}$). The resulting slope of this fit directly yields the measured numerical divergence angle $\theta$.

\begin{lstlisting}[caption={Extracting the Divergence Angle via Linear Regression}]
    # Select far-field propagation distances where linear divergence dominates
    valid_indices = np.abs(Z_m) > 0.5
    z_far = Z_m[valid_indices]
    sigma_far = sim_beam_sizes_m[valid_indices]
    
    # Perform a 1st-degree polynomial (linear) least-squares fit: 
    # sigma(z) = slope * z + intercept
    popt, cov = np.polyfit(z_far, sigma_far, 1, cov=True)
    
    # The slope (first element) corresponds to the numerical divergence angle
    theta_measured = popt[0] 
\end{lstlisting}

\begin{figure}[htbp]
    \centering
    \begin{minipage}[b]{0.45\textwidth}
        \centering
        \includegraphics[width=\textwidth]{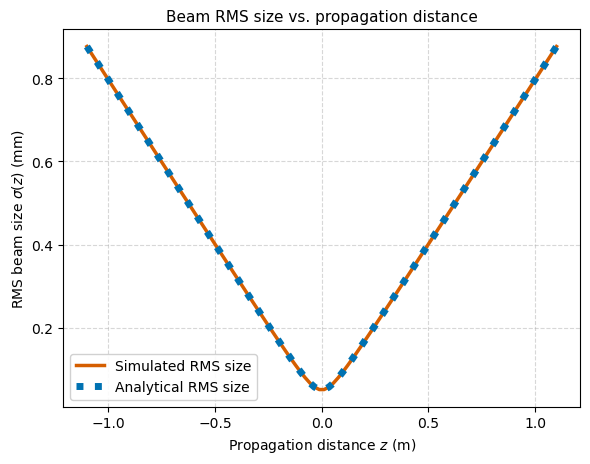}
        \caption{Comparison of simulated beam size with the analytical RMS beam size.}
        \label{fig:rms_sim_vs_ana}
    \end{minipage}
    \hfill
    \begin{minipage}[b]{0.45\textwidth}
        \centering
        \includegraphics[width=\textwidth]{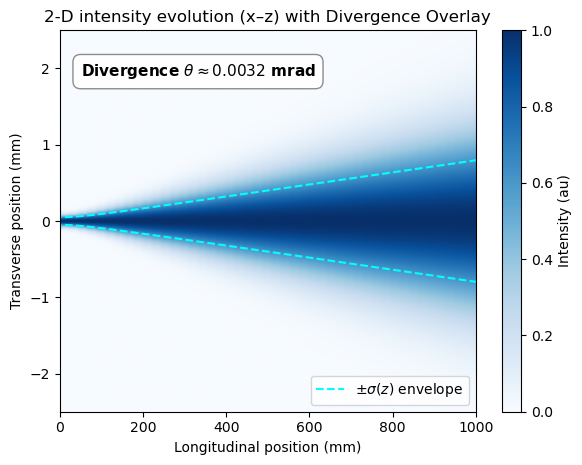}
        \caption{Beam propagation shown in two dimensions.}
        \label{fig:2Dintensity}
    \end{minipage}
\end{figure}

We overlay the analytical curve on the simulated data (Figure \ref{fig:rms_sim_vs_ana}). We can also plot the same beam propagation in 2D using the normalized intensity values of each axial slice (Figure \ref{fig:2Dintensity}).


\section{Modeling Optical Components}

A core application of computational Fourier optics is the numerical representation of physical optical elements. Following the linear systems framework, optical components are modeled as complex transmittance functions $t_A(x,y)$ that multiply the incident complex wavefront $U_{in}(x,y)$ to produce an exiting field $U_{out}(x,y) = t_A(x,y) U_{in}(x,y)$ immediately after the component. \cite{goodman1996, voelz2011} This section focuses on modeling a positive thin lens and analyzing the diffraction-limited focal evolution of a Gaussian beam.

\subsection{Theoretical Framework of the Thin Lens Transmittance}

A lens is classified as ``thin'' if a ray entering at coordinates $(x,y)$ on the input plane exits at approximately the same coordinates on the output plane, meaning the transverse displacement of light inside the lens is negligible \cite{goodman1996}. Under this paraxial approximation, the lens acts purely as a phase transformer that introduces a spatially varying phase delay proportional to the local thickness profile of the glass \cite{goodman1996, voelz2011}.

In general physical optical setups, a lens possesses a finite physical size that defines the pupil function $P(x,y)$. This pupil clips the spatial extent of the beam, limiting the spatial frequency bandwidth and establishing the effective diffraction-limited beam size at the focal plane \cite{goodman1996, voelz2011}. Neglecting constant phase delays, the general complex amplitude transmittance $t_l(x,y)$ of a positive lens of focal length $f$ is given by \cite{goodman1996}:
\begin{equation}
    t_l(x,y) = P(x,y) \exp\left(-\mathrm{j} \frac{k}{2f} (x^2 + y^2)\right)
\end{equation}
where $k = 2\pi/\lambda$ is the wave number. Physically, the quadratic phase term converts an incident planar wavefront into a converging spherical wavefront that focuses at a distance $z = f$ downstream. 

For our 1-D numerical simulations, we assume the incident beam is well contained within the lens boundary and is not clipped, allowing us to remove the pupil function ($P(x) \approx 1$). Consequently, the thin-lens phase mask simplifies to a function of the single transverse axis $x$ \cite{goodman1996}:
\begin{equation}
    L(x) = \exp\left(-\mathrm{j} \frac{\pi}{\lambda f} x^2\right)
\end{equation}

When explicit pupil clipping is included in broader physical models, we ensure that the physical aperture size is defined to be strictly smaller than the computational simulation window ($D_L < D_x$) so that edge-diffracted light remains fully contained within the grid. While full multi-element imaging systems are beyond the scope of this single-lens focal study, aperture pupil clipping plays a central role in imaging systems and represents a natural direction for future considerations of multi-lens systems \cite{goodman1996}.

The focal length $f$ of the system is governed by the thin lens equation (derived from the lensmaker's formula), which relates $f$ to the object (source) plane distance $p$ and conjugate image plane distance $q$ via the classical lens law \cite{goodman1996, voelz2011}:
\begin{equation}
    \frac{1}{f} = \frac{1}{p} + \frac{1}{q}
\end{equation}

\subsection{Numerical Implementation and Code Structure}

The physical principles of thin-lens wavefront modulation are implemented numerically by applying the quadratic lens phase mask onto the discretized spatial coordinate array. To visualize the complete longitudinal propagation through a lens system, transverse field slices are evaluated at discrete axial steps $z$ from the source plane ($z = -p_m$), through the lens plane ($z = 0$), to the conjugate image plane ($z = q_m$).

Because peak focal intensity increases by orders of magnitude at the waist compared to the unfocused wings, plotting raw unscaled intensity profiles directly can obscure the converging and diverging beam structure away from focus. To ensure the beam cone remains clearly visible across all axial positions, each transverse slice is normalized by its local peak intensity value \cite{voelz2011}:
\begin{equation}
    I_{\text{norm}}(x, z) = \frac{I(x, z)}{\max_x \{I(x, z)\}}
\end{equation}

\begin{figure}[htbp]
    \centering
    \includegraphics[width=0.75\textwidth]{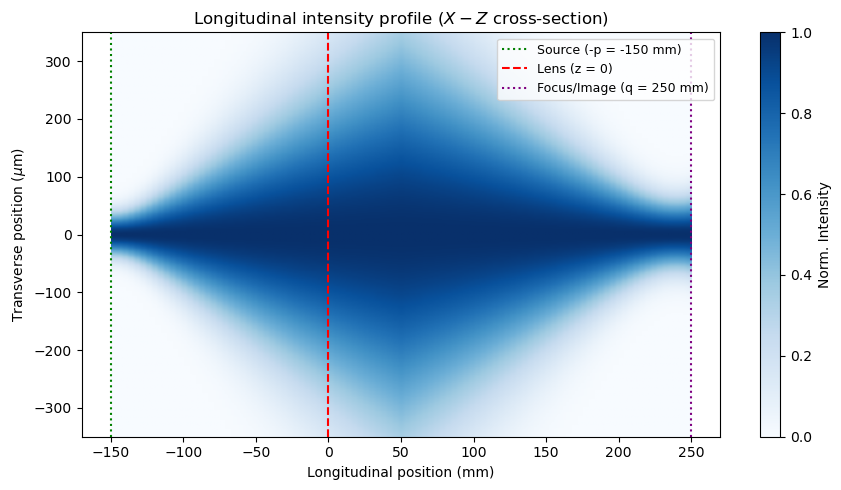}
    \caption{Longitudinal intensity cross-section ($X-Z$ map) showing coherent beam convergence through a positive thin lens at $z = 0\text{ mm}$. The beam originates at the source plane $z = -p_m = -150\text{ mm}$ and focuses at the conjugate image plane $z = q_m = +250\text{ mm}$. Each transverse slice is normalized to its local peak intensity to render the light cone clearly visible across all axial positions \cite{goodman1996, voelz2011}.}
    \label{fig:longitudinal_map}
\end{figure}

Listing \ref{lst:propagation} details the Python implementation for constructing the thin-lens phase mask, propagating the field to the conjugate focal plane, and accumulating the normalized $X-Z$ longitudinal cross-sectional map shown in Figure \ref{fig:longitudinal_map}.

\begingroup
\centering
\begin{lstlisting}[language=Python, caption={Numerical Implementation of Lens Propagation and Longitudinal Mapping}, label={lst:propagation}]

import numpy as np
import matplotlib.pyplot as plt

# Physical parameters
p_m = 0.15                  # Object/source distance to lens (m)
q_m = 0.25                  # Lens to conjugate image plane distance (m)
wavelength_m = 500e-9       # Source wavelength (m)

# Thin Lens Law calculation of focal length
f_m = 1.0 / (1.0 / p_m + 1.0 / q_m) 

# Quadratic lens phase mask in 1-D
lens_mask = np.exp(-1j * (2 * np.pi / wavelength_m) / (2 * f_m) * x_m**2)

# Field immediately after the lens (Ep is the propagated source field from z = -p_m)
E_after_lens = Ep * lens_mask

# Propagate complex field from lens plane (z = 0) to image plane (z = q_m)
Ef = propTF(E_after_lens, Dx_m, wavelength_m, q_m)

# Longitudinal scan: accumulate normalized transverse slices across Z
z_scan = np.linspace(-p_m, q_m, 200)
I_zx = []

for z in z_scan:
    if z &lt; 0:
        # Propagation before lens (from source plane at z = -p_m to z)
        Ez = propTF(E0, Dx_m, wavelength_m, z + p_m)
    else:
        # Propagation after lens (from lens plane at z = 0 to z)
        Ez = propTF(E_after_lens, Dx_m, wavelength_m, z)
    
    Iz = np.abs(Ez)**2
    I_zx.append(Iz / np.max(Iz)) # Normalize each slice so beam remains visible

I_zx = np.array(I_zx)
\end{lstlisting}
\endgroup

\subsection{Sampling Constraints and Transverse Magnification Scaling}

A major requirement in numerical wave-optics simulations is ensuring that both free-space propagations and thin-lens phase modulations are sampled without spatial phase aliasing \cite{voelz2011}. For a propagation distance $z$ across a transverse grid or aperture window of size $D$, the spatial sampling interval $\Delta x$ must satisfy the fundamental sampling criterion \cite{voelz2011}:
\begin{equation}
    \Delta x \le \frac{\lambda z}{D}
\end{equation}

While general free-space propagation limits dictate the minimum grid resolution required to prevent boundary aliasing over distance $z$, focusing through a lens specifically introduces constraints governed by transverse magnification $M_t$ \cite{goodman1996, voelz2011}.

According to paraxial geometric and wave optics, the transverse magnification $M_t$ for a single-lens system with object distance $p$ and conjugate image distance $q$ is defined by the distance ratio \cite{goodman1996}:
\begin{equation}
    M_t = -\frac{q}{p}
\end{equation}
When the input source distance $p$ and initial waist size $\sigma_0$ are held constant, the focused RMS spot size $\sigma_{\text{image}}$ at the conjugate plane $z = q$ scales linearly with the throw distance $q$:
\begin{equation}
    \sigma_{\text{image}}(q) = \sigma_0 \cdot |M_t| = \sigma_0 \left( \frac{q}{p} \right)
\end{equation}

This magnification scaling governs two complementary sampling considerations during numerical focal simulations:
\begin{enumerate}
    \item \textbf{Focal Waist Resolution (Short $q$):} At short image distances $q$ (small magnification $|M_t| < 1$), the beam converges to a tight, highly localized focal waist. The spatial grid spacing $\Delta x$ must be sufficiently fine ($\Delta x \ll \sigma_{\text{image}}$) to resolve the focal peak without under-sampling or corrupting numerical moment calculations \cite{voelz2011}.
    \item \textbf{Window Boundary Expansion (Large $q$):} At large image distances $q$ (large magnification $|M_t| > 1$), the expanding beam beyond focus requires a sufficiently wide spatial window $D_x = M \Delta x$ to prevent light from spilling over array boundaries and producing periodic wrap-around artifacts \cite{voelz2011}.
\end{enumerate}

\subsection{Focused Spot Size vs. Image Distance}

To quantitatively verify this magnification model, we simulate focal spot convergence across a sweep of image distances $q_m \in [0.05\text{ m}, 0.50\text{ m}]$ for a fixed object distance $p_m = 0.15\text{ m}$. For each step, the focal length $f_m$ is dynamically updated using the thin lens law ($1/f_m = 1/p_m + 1/q_m$) to maintain focus at the target observation plane. 

\begin{figure}[htbp]
    \centering
    \begin{minipage}{0.99\textwidth}
        \centering
        \includegraphics[width=\textwidth]{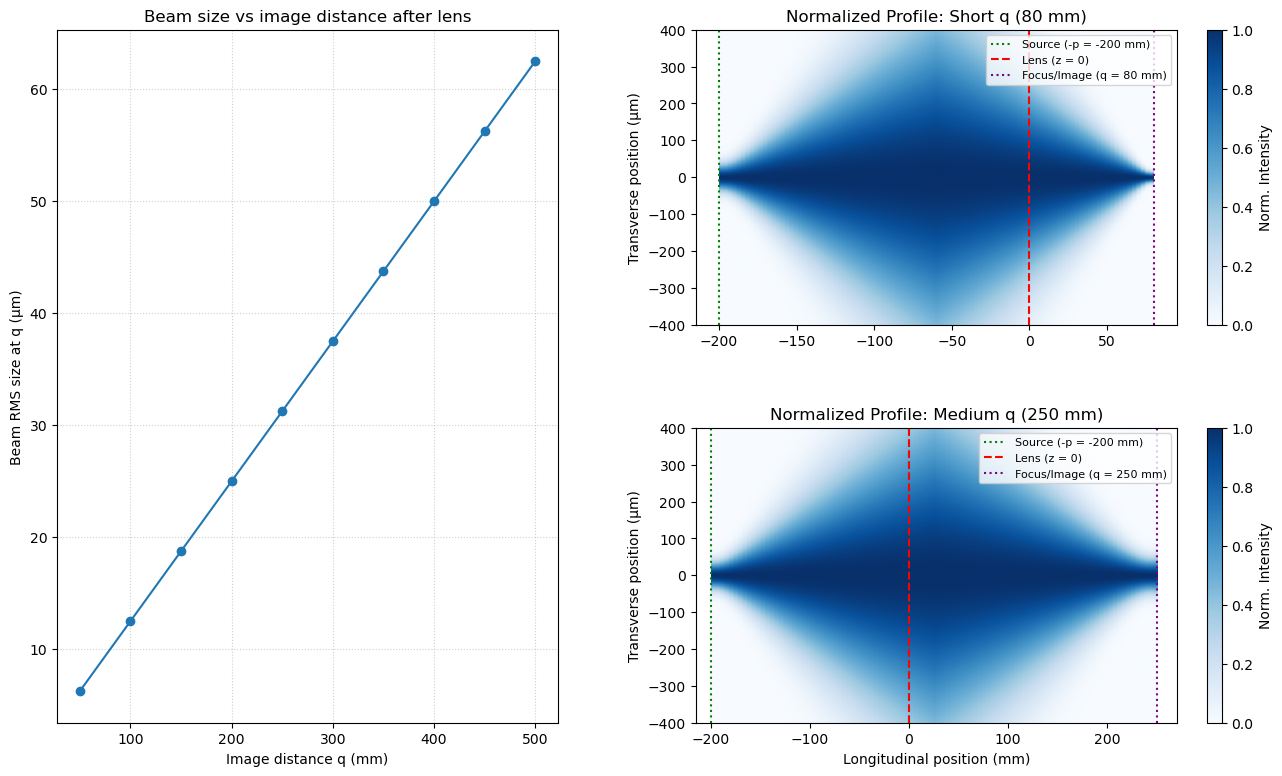}
        \caption{Quantitative magnification analysis of thin-lens focusing: \textbf{(a)} Focused RMS spot size $\sigma_{\text{image}}$ as a function of conjugate image distance $q_m$ for $p_m = 0.15\text{ m}$, verifying the linear relationship $|M_t| = q_m / p_m$; \textbf{(b, c)} Stacked 2-D transverse focal intensity distributions ($I(x,y)$) comparing the tight, high-contrast spot at short throw ($q_m = 0.10\text{ m}$) against the magnified profile at intermediate throw ($q_m = 0.25\text{ m}$).}
        \label{fig:beamsize}
    \end{minipage}
\end{figure}

Figure \ref{fig:beamsize}a plots the resulting focused RMS beam radius $\sigma_{\text{image}}$ as a function of $q_m$, demonstrating an exact linear agreement with the analytical magnification scaling $|M_t| = q_m / p_m$. 

To illustrate the spatial impact of magnification on the transverse profile, Figures \ref{fig:beamsize}b and \ref{fig:beamsize}c present 2-D cross-sectional intensity distributions ($I(x,y)$) evaluated at two distinct conjugate image planes:
\begin{itemize}
    \item \textbf{Short Throw Distance ($q_m = 0.10\text{ m}$, $|M_t| \approx 0.67$):} At short image distances, the beam undergoes strong convergence, concentrating the optical power into a tight, high-contrast focal spot with a compact spatial footprint (Figure \ref{fig:beamsize}b).
    \item \textbf{Intermediate Throw Distance ($q_m = 0.25\text{ m}$, $|M_t| \approx 1.67$):} At larger throw distances, the increased transverse magnification spreads the optical power across a wider spatial area, yielding a broader spot profile (Figure \ref{fig:beamsize}c).
\end{itemize}

It is important to note that the linear magnification ratio holds strictly under ideal conditions where \textit{no pupil or aperture clipping} occurs. When physical aperture limits or severe pupil clipping are introduced, truncated optical power redistribution causes the actual imaged spot to appear larger than predicted by simple geometric scaling. Throughout this analysis, we assume an idealized numerical aperture window without truncation artifacts—a constraint that warrants further investigation regarding finite lens apertures in future work.

Listing \ref{lst:spot_size_simulation} provides the Python code used to execute the image-distance sweep, calculate the focal waist sizes, and generate the stacked 2-D intensity maps.

\begin{lstlisting}[language=Python, caption={Parametric Study of Beam Size vs. Image Distance and 2-D Focal Spot Profiles}, label={lst:spot_size_simulation}]
import numpy as np
import matplotlib.pyplot as plt

# Fixed simulation parameters
p_m = 0.15                 # Fixed object distance (150 mm)
wavelength_m = 500e-9      # Source wavelength (500 nm)

# Sweep of conjugate image distances q_m
q_sweep_m = np.linspace(0.05, 0.50, 10)
sigma_image_m = []

# Loop over q distances to compute focal spot sizes
for q_m in q_sweep_m:
    f_m = 1.0 / (1.0 / p_m + 1.0 / q_m) # Focal length via Thin Lens Law
    lens_mask = np.exp(-1j * (np.pi / (wavelength_m * f_m)) * x_m**2)
    
    E_after_lens = Ep * lens_mask
    Ef = propTF(E_after_lens, Dx_m, wavelength_m, q_m)
    If = np.abs(Ef)**2
    
    # Calculate numerical RMS waist size from second moment
    x_mean = np.sum(x_m * If) / np.sum(If)
    sig_z = np.sqrt(np.sum((x_m - x_mean)**2 * If) / np.sum(If))
    sigma_image_m.append(sig_z * 1e6) # Convert to micrometers

\end{lstlisting}

\subsection{Quantitative Verification of System Magnification}

To validate the numerical accuracy of our wave-optics framework, we perform a statistical characterization of the beam size at both the source plane ($z = -p_m$) and conjugate focal plane ($z = q_m$) using the Root Mean Square (RMS) width. The RMS width provides a robust measure of the spatial extent of the intensity distribution $I(x)$ by calculating its second central moment \cite{voelz2011}:
\begin{equation}
    \sigma_{\text{rms}} = \sqrt{\frac{\sum (x - \bar{x})^2 I(x)}{\sum I(x)}}
\end{equation}
where $\bar{x} = \frac{\sum x I(x)}{\sum I(x)}$ is the centroid (intensity-weighted first moment) of the beam, and $\sum I(x)$ provides power normalization, ensuring that $\sigma_{\text{rms}}$ is invariant to overall intensity scaling \cite{voelz2011}.

Listing \ref{lst:magnification_verification} tracks the beam width from the initial object plane to the focused image plane for $p_m = 0.15\text{ m}$ and $q_m = 0.25\text{ m}$.

\begin{lstlisting}[language=Python, caption={Numerical Verification of Beam Magnification}, label={lst:magnification_verification}]
# Compute field at conjugate focus (q_m = 0.25 m)
Ep_focus = propTF(E_after_lens, Dx_m, wavelength_m, q_m)
I_focus = np.abs(Ep_focus)**2

# RMS width of initial beam at object plane (source)
x_mean_init = np.sum(x_m * np.abs(E0)**2) / np.sum(np.abs(E0)**2)
x_std_init = np.sqrt(np.sum((x_m - x_mean_init)**2 * np.abs(E0)**2) / np.sum(np.abs(E0)**2))

# RMS width of focused beam at conjugate image plane (focus)
x_mean_focus = np.sum(x_m * I_focus) / np.sum(I_focus)
x_std_focus = np.sqrt(np.sum((x_m - x_mean_focus)**2 * I_focus) / np.sum(I_focus))
\end{lstlisting}

Executing this script yields:
\begin{verbatim}
Initial RMS beam size: 25.00 um
Focused RMS beam size: 41.67 um
Calculated Ratio (initial/focused): 0.60
\end{verbatim}

\subsubsection{Interpretation of Results}

The transverse magnification $M_t$ predicted by geometric optics for $p_m = 0.15\text{ m}$ and $q_m = 0.25\text{ m}$ is $|M_t| = q_m / p_m = 0.25 / 0.15 \approx 1.667$. Applying this scale factor to our initial Gaussian beam waist ($\sigma_{\text{object}} = 25.00\ \mu\text{m}$) predicts a focused spot size of exactly:
\begin{equation}
    \sigma_{\text{image}} = \sigma_{\text{object}} \cdot |M_t| = 25.00\ \mu\text{m} \times 1.667 = 41.67\ \mu\text{m}
\end{equation}

Our wave-optics simulation—computed via the \texttt{propTF} function propagator—recovers this exact value of $41.67\ \mu\text{m}$. The calculated scaling ratio of $0.60$ represents the inverse magnification ($1/|M_t| = p_m / q_m$). This perfect numerical agreement verifies that our discrete Fourier framework preserves physical spatial scaling, energy conservation, and imaging magnification laws \cite{goodman1996, voelz2011}.


\section{Simulation of Diffractive Elements and Gratings}

A key area of wave-optics modeling is the simulation of diffractive optical elements, which manipulate wavefronts through localized diffraction rather than refraction \cite{goodman1996}. In the linear systems framework, a diffracting screen is represented by its complex amplitude transmittance function $t_A(x,y)$, defined as the ratio of the exiting complex field to the incident complex field immediately behind the screen \cite{goodman1996, voelz2011}:
\begin{equation}
    E_{\text{exiting}}(x,y) = E_{\text{incident}}(x,y) \cdot t_A(x,y)
\end{equation}

Diffractive elements are generally categorized into amplitude gratings, which control the magnitude of the field, and phase gratings, which alter the local optical path length without weakening the field \cite{goodman1996, voelz2011}. This section focuses on modeling a binary amplitude grating and analyzing the resulting spatial frequency separation as the field propagates.

\subsection{Binary Amplitude Grating Formulation and Sampling}

A binary diffraction grating consists of a periodic array of opaque and transparent slits acting as a square-wave transmittance mask $t_A(x)$ with a period $d$ that alternates between $0$ (absorption) and $1$ (complete transmission) \cite{goodman1996}. 

To model this square-wave profile directly, we take the signum function of a sine wave and scale/shift it to make the transmittance strictly positive, alternating cleanly between $0$ and $1$:

\begin{equation}
    t_A(x) = \frac{1}{2} \left[ 1 + \text{sgn}\left(\sin\left(\frac{2\pi}{d} x\right)\right) \right]
\end{equation}

We implement this periodic transmittance in a 1-D spatial grid. The number of discrete grid points $M_d$ that span a single grating period $d$ must satisfy \cite{voelz2011}:
\begin{equation}
    M_d = \frac{d}{\Delta x} \ge 2
\end{equation}
to avoid spatial aliasing. While a minimum of $2$ samples satisfies Eq 28, a larger value of $M_d$ is required in practice to represent the abrupt, step-like binary phase and amplitude boundaries without introducing numerical ringing artifacts \cite{voelz2011, goodman1996}.

With a grid spacing of $\Delta x = 1\ \mu\text{m}$ and a grating period of $d = 10\ \mu\text{m}$, our grid allocates exactly $M_d = 10$ samples per cycle, ensuring a highly resolved representation of the binary transition boundaries.

\begin{lstlisting}[language=Python, caption={Generating a Binary Grating Mask}, label={lst:binary_grating}]
period_m = 10e-6 # Grating period (10 micrometers)

# Create a binary mask where transmission is 1 (True) or 0 (False)
grating = 0.5 * (1 + np.sign(np.sin(2 * np.pi * x_m / period_m)))

# Apply the transmittance function to the propagated beam
Eg = Ep * grating 
Ig = np.abs(Eg)**2
\end{lstlisting}

\begin{figure}[htbp]
    \centering
    \begin{minipage}{0.5\textwidth}
        \centering
        \includegraphics[width=\textwidth]{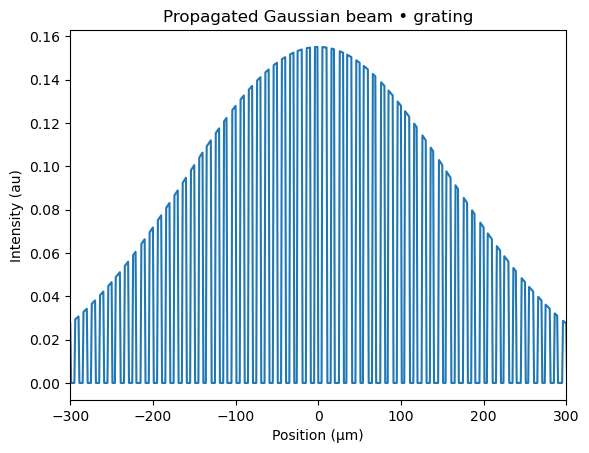}
        \caption{Transverse intensity distribution ($I_g = |E_g|^2$) of a Gaussian beam immediately after passing through the binary grating mask, showing the modulation of the Gaussian envelope.}
        \label{fig:1Dgrating}
    \end{minipage}
\end{figure}

When the propagated Gaussian field $E_p$ interacts with this element, the exiting complex field $E_g$ is the point-wise product of the wave and the mask. This results in an altered near-field intensity distribution where the smooth Gaussian envelope is masked by the periodic structure, as shown in Figure \ref{fig:1Dgrating}.

\subsection{Tilt in Fourier Propagation}

Beyond diffraction from periodic structures, wave-front manipulation can also be introduced via phase ramps, commonly referred to as tilt. Applying a linear phase modulation across the initial field tilts its direction of travel, causing the entire beam to steer off-axis during propagation.

An angular tilt $\theta_{\text{tilt}}$ modifies the optical field such that, under the paraxial approximation, it translates to a linear transverse displacement over propagation distance $z$ \cite{goodman1996}:
\begin{equation}
    \Delta x(z) \approx z \tan\theta_{\text{tilt}} \approx z \theta_{\text{tilt}}
\end{equation}

For a specified tilt angle of $\theta_{\text{tilt}} = 10\text{ mrad}$ ($0.01\text{ rad}$), the expected transverse displacement at a propagation distance of $z_p = 100\text{ mm}$ is:
\begin{equation}
    \Delta x(100\text{ mm}) \approx 100\text{ mm} \times 0.01\text{ rad} = 1\text{ mm} \quad (1000\ \mu\text{m})
\end{equation}

To track this trajectory, we simulate the propagation loop by applying the tilt factor to the initial field $E_0$ and evaluating the transverse intensity profile across axial steps up to $z_p = 100\text{ mm}$.

\begin{lstlisting}[language=Python, caption={Longitudinal Propagation of a Tilted Gaussian Beam}, label={lst:tilt_propagation}]
zp_m = 100e-3 # propagation distance, m

Et = E0 * tilt # apply tilt to the initial field
Ept = propTF(Et, Dx_m, wavelength_m, zp_m) # propagate
Ipt = np.abs(Ept)**2 # intensity

# Longitudinal grid setup
num_pts = 100
z_after = np.linspace(0, zp_m, num_pts)

# Propagation loop with per-slice peak normalization
I_zx = []
for z in z_after:
    Ep = propTF(Et, Dx_m, wavelength_m, z)
    I_slice = np.abs(Ep) ** 2

    # Normalize slice to its peak value
    max_val = np.max(I_slice)
    if max_val > 0:
        I_slice = I_slice / max_val

    I_zx.append(I_slice)

I_zx = np.array(I_zx)
\end{lstlisting}

\begin{figure}[htbp]
    \centering
    \begin{minipage}{0.7\textwidth}
        \centering
        \includegraphics[width=\textwidth]{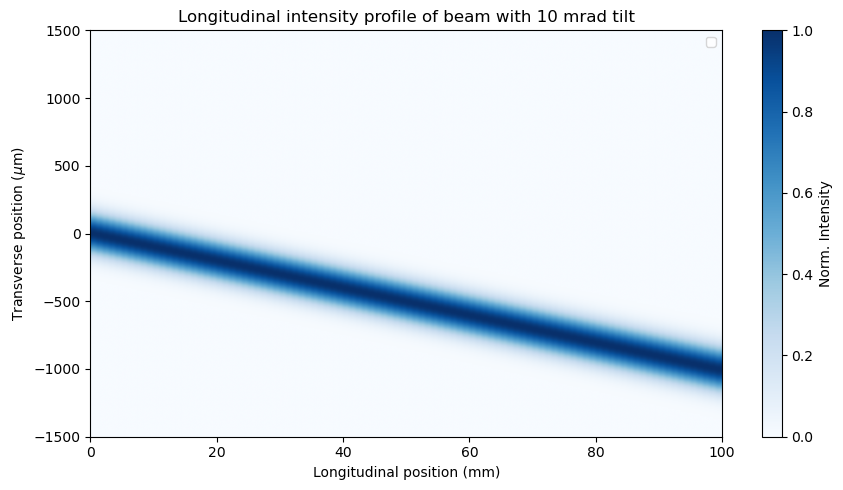}
        \caption{Longitudinal intensity profile ($X-Z$ cross-section) of a Gaussian beam with a $10\text{ mrad}$ wavefront tilt, showing linear transverse steering over a $100\text{ mm}$ propagation range.}
        \label{fig:tilt_beam}
    \end{minipage}
\end{figure}

The resulting longitudinal intensity profile, shown in Figure \ref{fig:tilt_beam}, clearly illustrates the steering effect. As the beam propagates from $z = 0$ to $z = 100\text{ mm}$, its center shifts continuously downward to approximately $-1000\ \mu\text{m}$ ($1\text{ mm}$). This exact agreement with analytical predictions confirms that the transfer function propagator correctly handles directional shifts associated with off-axis beam propagation \cite{goodman1996, voelz2011}.

\subsection{Longitudinal Propagation and Angular Separation}

We can map the 2-D longitudinal evolution of the diffracted field across the $X-Z$ plane by repeating the propagation calculations over a sequence of axial steps $z$. As light travels away from the grating plane, the individual Fourier components (the diffracted orders) begin to separate.

According to classical grating diffraction theory, the diffracted orders exit the screen at angles $\theta_n$ governed by the grating equation \cite{goodman1996}:
\begin{equation}
    \sin\theta_n = n \frac{\lambda}{d}
\end{equation}
Under the paraxial approximation, these angular deflections translate to a linear spatial separation $\Delta x_n$ along the transverse axis as a function of propagation distance $z$ \cite{goodman1996}:
\begin{equation}
    \Delta x_n(z) \approx z \sin\theta_n = n \frac{\lambda z}{d}
\end{equation}

For our simulation parameters ($\lambda = 500\text{ nm}$, $d = 10\ \mu\text{m}$), the first-order deflection angle is:
\begin{equation}
    \theta_{\pm 1} \approx \pm \frac{500 \times 10^{-9}\text{ m}}{10 \times 10^{-6}\text{ m}} = \pm 0.05\text{ rad} \quad (\pm 50\text{ mrad})
\end{equation}
At a propagation distance of $z = 100\text{ mm}$ ($0.1\text{ m}$), the expected transverse separation of the $+1$ and $-1$ diffracted orders from the central optical axis is:
\begin{equation}
    \Delta x_{\pm 1} \approx \pm 0.1\text{ m} \times 0.05\text{ rad} = \pm 5\text{ mm}
\end{equation}

In the scanning loop used to construct this longitudinal propagation map, for each step $z$, we propagate the field $E_g$, normalize its intensity to maintain visual contrast, and calculate the evolving beam RMS width to monitor divergence.

\begin{lstlisting}[language=Python, caption={Binary Grating Mask in Two Dimensions}, label={lst:longitudinal_grating}]
# Define axial steps from z = 0 to z = 0.5 m
Z = np.linspace(0, 0.5, 500)  
sim_beam_sizes = []

Ixz = np.zeros((len(Z), len(x_m)))
for i, z in enumerate(Z):
    Ed = propTF(Eg, Dx_m, wavelength_m, z)
    Id = np.abs(Ed)**2
    Ixz[i, :] = Id / np.max(Id)
    
    x_mean = np.sum(x_m * Id) / np.sum(Id)
    x_std = np.sqrt(np.sum((x_m - x_mean)**2 * Id) / np.sum(Id))
    sim_beam_sizes.append(x_std)

sim_beam_sizes = np.array(sim_beam_sizes)



\end{lstlisting}


\begin{figure}[htbp]
    \centering
    \begin{minipage}{0.7\textwidth}
        \centering
        \includegraphics[width=\textwidth]{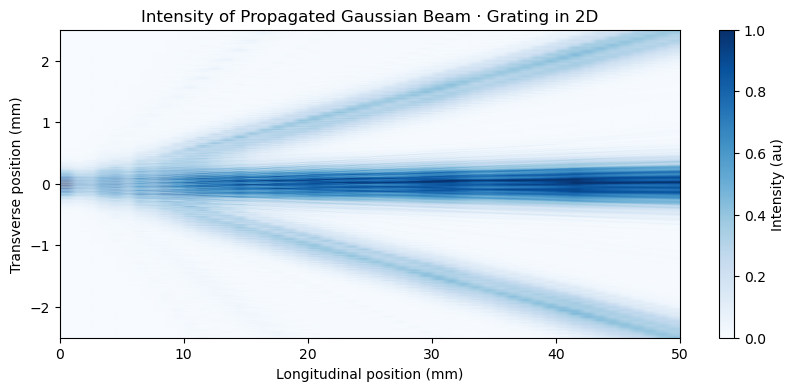}
        \caption{Longitudinal intensity profile ($X-Z$ cross-section) of a Gaussian beam after passing through the grating, illustrating the spatial fan-out of the $0$ and $\pm 1$ diffraction orders.}
        \label{fig:2Dgrating}
    \end{minipage}
\end{figure}

The resulting cross-sectional intensity map, shown in Figure \ref{fig:2Dgrating}, visually illustrates this diffractive branching. Near the grating plane ($z = 0$), the field remains confined. As $z$ increases, the wave-front splits into distinct fan-out cones. The central branch represents the $0$-order beam containing 25\% of the incident power, while the branching arms represent the $+1$ and $-1$ diffracted orders, each containing 6.25\% of the incident power. 


The spatial separation observed in Figure \ref{fig:2Dgrating} matches our analytical paraxial calculation of $\pm 5\text{ mm}$ at $z = 100\text{ mm}$, demonstrating that the paraxial propagator \texttt{propTF} correctly preserves the wave scaling, angular dispersion, and physical scaling limits of Fourier optics \cite{goodman1996, voelz2011}.

\section{Experimental Characterization of Diffractive Focusing Elements}

While refractive optics dominate visible-light imaging, diffractive optical elements such as Fresnel zone plates are essential in regions of the electromagnetic spectrum—such as the extreme ultraviolet and X-ray regimes—where conventional optical glasses are highly absorbing \cite{miyakawa2014}. FZPs manipulate wavefronts through localized diffraction rather than refraction, focusing light by constructive interference of waves emanating from concentric circular zones. This section details the experimental characterization of a Fresnel zone plate and compares the measured focal evolution against paraxial wave propagation theory.

\subsection{Theoretical Framework of Fresnel Zone Plates and Digital Holography}

A Fresnel zone plate (FZP) is a diffractive lens consisting of concentric alternating transparent and opaque zones. For a rigorous treatment of zone plate transmittance functions, multi-order focal positions ($f_n = \pm f/n$), and diffractive efficiencies, the reader is referred to Atwood \cite{attwood2017}. 

To maximize diffraction efficiency in computational optics, we implement a phase-only binary holographic reconstruction workflow. 

\subsubsection{Numerical Implementation and Code Structure}

The simulation isolates the phase components of the forward-propagated source field ($E_{\text{source\_prop}}$) and back-propagated image field ($E_{\text{image\_prop}}$) to construct a stationary interference pattern without amplitude attenuation \cite{goodman1996,voelz2011}. Hard thresholding is then applied to generate a binary mask suitable for digital wavefront reconstruction.

Listing \ref{lst:holographic_reconstruction} consolidates this workflow: isolating the phase superposition, binarizing to form the mask, modulating with a reference plane wave, and back-propagating to the focal plane.

\begin{lstlisting}[language=Python, caption={Numerical Recombination, Binarization, and Digital Holographic Reconstruction}, label={lst:holographic_reconstruction}]
# 1. Isolate phase and compute phase-only superposition
I_sum = np.abs(E_source_prop + E_image_prop)**2
phase_source = np.exp(1j * np.angle(E_source_prop))
phase_image = np.exp(1j * np.angle(E_image_prop))
I_sum2 = np.abs(phase_source + phase_image)

# 2. Threshold to generate binary hologram mask (0 or 1)
hologram = (I_sum2 / 2) >= 0.5

# 3. Wavefront Playback: Use a uniform incident reference beam (plane wave)
E_source_holo = np.ones(hologram.shape)

# 4. Modulate reference wave by the binary hologram
E_holo = E_source_holo * hologram

# 5. Propagate back to focal plane (-z_prop_m to focus down to a point)
E_image_holo = propTF(E_holo, Dx_m, 500e-9, -z_prop_m)
\end{lstlisting}

\begin{figure}[htbp]
    \centering
    \includegraphics[width=0.9\textwidth]{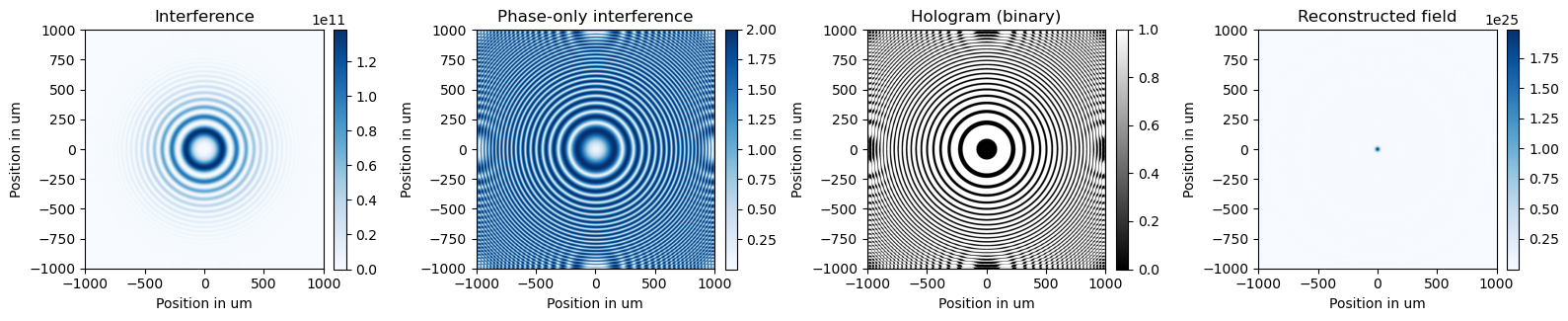}
    \caption{Simulation and reconstruction of holographic interference patterns: (a) Total continuous intensity distribution $I_{\text{sum}}$, (b) Phase-only superposition $I_{\text{sum2}}$, (c) Binarized digital hologram mask, and (d) Reconstructed focal intensity profile $|E_{\text{image\_holo}}|^2$. Spatial dimensions are in micrometers ($\mu\text{m}$).}
    \label{fig:hologram}
\end{figure}

\subsubsection{Physical Interpretation of the Reconstructed Field}

As illustrated in Figure~\ref{fig:hologram}, the workflow progresses through four physical regimes:
\begin{enumerate}
    \item \textbf{Continuous Interference (Fig.~\ref{fig:hologram}a):} Quadratic phase curvature mismatch between the source and image fields produces high-contrast concentric interference fringes with a central destructive intensity null \cite{goodman1996, voelz2011}.
    \item \textbf{Phase-Only Superposition (Fig.~\ref{fig:hologram}b):} Setting local amplitudes to unity isolates the phase carrier, maintaining 100\% modulation contrast and preventing high-frequency outer fringes from decaying into noise.
    \item \textbf{Binary Hologram Mask (Fig.~\ref{fig:hologram}c):} Thresholding at $0.5$ converts the continuous phase pattern into a binary transmission grating analogous to a binary FZP.
    \item \textbf{Reconstructed Focus (Fig.~\ref{fig:hologram}d):} During wavefront playback, the binarized hologram is illuminated by a replica of the original source reference wave ($E_{\text{source\_holo}}$) and propagated to the conjugate focal plane. Despite the complete loss of amplitude information and the severe non-linear binarization of the phase carrier, the reconstructed intensity distribution $|E_{\text{image\_holo}}|^2$ successfully focuses into a highly localized, diffraction-limited central spot. 
\end{enumerate}

\section{Data Collection}
\subsection{Intro to Data Collection}
To validate our numerical transfer-function propagator (\texttt{propTF}) against physical wave fields, we constructed an optical test bench to perform camera-based through-focus profiling. Before presenting the reconstructed longitudinal intensity distributions, it is essential to establish the physical geometry, component specifications, and data collection routine of the apparatus.


\begin{figure}[htbp]
    \centering
    \begin{minipage}{0.85\textwidth}
        \centering
        \includegraphics[width=\textwidth]{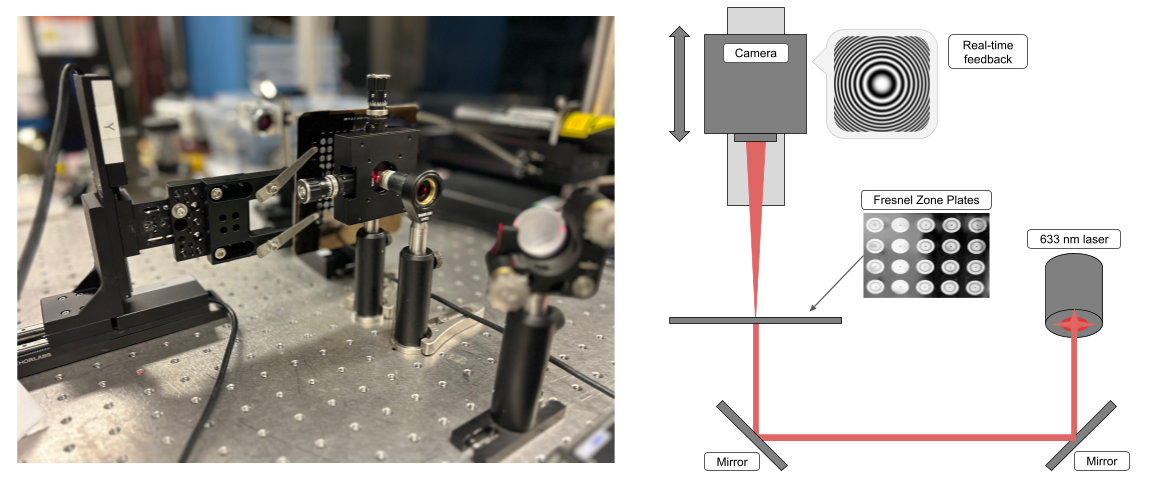}
        \caption{Photo and schematic diagram of the optical bench setup for through-focus spatial profiling: A spatially filtered, collimated laser beam ($\lambda = 633\text{ nm}$) illuminates the test optic (a Fresnel zone plate). A high-resolution camera profiler ($3.45\ \mu\text{m}$ pixel pitch) is mounted on a linear translation stage and swept along the optical axis ($z$-axis) across the focal volume, capturing a sequence of 2-D transverse intensity distributions $I(x,y; z)$ at discrete axial positions.}
        \label{fig:experimental_setup}
    \end{minipage}
\end{figure}

As illustrated in the schematic diagram of Figure \ref{fig:experimental_setup}, the experimental apparatus consists of three primary optical sections:

\begin{enumerate}
    \item \textbf{Coherent Illumination Source:} A spatially filtered and collimated laser source ($\lambda = 633\text{ nm}$) delivers a uniform, clean planar wavefront across the clear aperture of the optical bench.
    \item \textbf{Test Optical Element:} The collimated beam interacts with the test optic—a Fresnel zone plate (FZP)—which impresses a converging quadratic phase delay onto the incident wavefront to form a diffraction-limited focus.
    \item \textbf{Axial Camera Profiling Assembly ($z$-Scan):} A high-resolution camera profiler (pixel pitch $\Delta x_{\text{pixel}} = 3.45\ \mu\text{m}$, 8-bit ADC, full well capacity $10,000\ e^-$) is mounted on an axial linear translation stage aligned along the optical axis ($z$).
\end{enumerate}

\subsection{Radiometric Calibration to Physical Power Density}

To convert raw camera digital numbers (ADU) into absolute physical irradiance ($\text{W/m}^2$), we implement a multi-stage radiometric conversion pipeline.

First, raw 8-bit (sensor-specific) digital readings $I_{\text{ADU}}$ are converted to photoelectron counts $I_{e^-}$ using the sensor full well capacity ($\text{FWD} = 10,000\ e^-$):
\begin{equation}
    I_{e^-} = I_{\text{ADU}} \times \left( \frac{\text{FWD}}{2^{\text{bit\_depth}} - 1} \right)
\end{equation}

Accounting for the camera quantum efficiency ($\text{QE} = 0.8$), the photoelectron field is converted into incident photon counts $I_{\text{photon}} = I_{e^-} / \text{QE}$. The optical energy $E_{\text{pixel}}$ deposited per pixel is calculated using Planck's photon energy relation $E_{\gamma} = hc/\lambda$ at $\lambda = 633\text{ nm}$ (source specific) ($E_{\gamma} \approx 3.14 \times 10^{-19}\text{ J}$):
\begin{equation}
    E_{\text{pixel}} = I_{\text{photon}} \cdot \left( \frac{hc}{\lambda} \right)
\end{equation}

Dividing by the camera exposure time used ($\tau = 75\ \mu\text{s}$) converts energy to optical power $P_{\text{pixel}} = E_{\text{pixel}} / \tau$ in Watts. Finally, dividing by the individual pixel area ($A_{\text{pixel}} = \Delta x_{\text{pixel}}^2$, with pixel pitch $\Delta x_{\text{pixel}} = 3.45\ \mu\text{m}$) yields the absolute power density distribution $I_{\text{physical}}(x,y)$ in $\text{W/m}^2$:
\begin{equation}
    I_{\text{physical}}(x,y) = \frac{P_{\text{pixel}}}{\Delta x_{\text{pixel}}^2}
\end{equation}

Listing \ref{lst:radiometric_scaling} details the implementation of this calibration routine.

\begin{lstlisting}[language=Python, caption={Converting Detector ADU Readings to Physical Irradiance (W/m$^2$)}, label={lst:radiometric_scaling}]
# Sensor specifications
fwd = 10000              # Full well depth (electrons)
bit_depth = 8            # 8-bit ADC (0 to 255)
qe = 0.8                 # Quantum efficiency
exposure_time_s = 75e-6  # Exposure time (75 microseconds)
pixel_size_m = 3.45e-6   # Pixel size (3.45 micrometers)
wavelength_m = 633e-9    # Laser wavelength (633 nm)

# Physical constants
h_Js = 6.62607015e-34    # Planck's constant (J*s)
c_ms = 3e8              # Speed of light (m/s)

# 1. ADU -> Photoelectrons
electrons_per_adu = fwd / (2**bit_depth - 1)
img_electron = data * electrons_per_adu

# 2. Photoelectrons -> Incident Photons
img_photon = img_electron / qe

# 3. Photons -> Deposited Energy (Joules)
E_J = h_Js * c_ms / wavelength_m
img_J = img_photon * E_J

# 4. Energy -> Power (Watts)
img_power_W = img_J / exposure_time_s

# 5. Power -> Power Density (W/m^2)
img_power_density_W_m2 = img_power_W / (pixel_size_m**2)

# Spatial axes in millimeters
x_mm = np.linspace(0, img_power_density_W_m2.shape[5] * pixel_size_m * 1e3, img_power_density_W_m2.shape[5])
y_mm = np.linspace(0, img_power_density_W_m2.shape * pixel_size_m * 1e3, img_power_density_W_m2.shape)

\end{lstlisting}

\subsection{Experimental Spatial Profiling and Data Extraction}

To experimentally map the diffractive focusing behavior, a through-focus series is recorded by capturing the transverse spatial profile of the beam at discrete axial planes along the optical axis. The experimental run consists of a 21-step scan spanning a total range of $2.0\text{ m}$ around the nominal focal volume of an aberration-free test zone plate. 

The captured 2-D images are integrated along the vertical axis to compress the spatial data into 1-D transverse cross-sections. These 1-D profile slices are normalized to prevent saturation and then concatenated sequentially to map the longitudinal intensity evolution $I(x,z)$ across the focal volume.


\begin{figure}[htbp]
    \centering
    \begin{minipage}{0.9\textwidth}
        \centering
        \includegraphics[width=\textwidth]{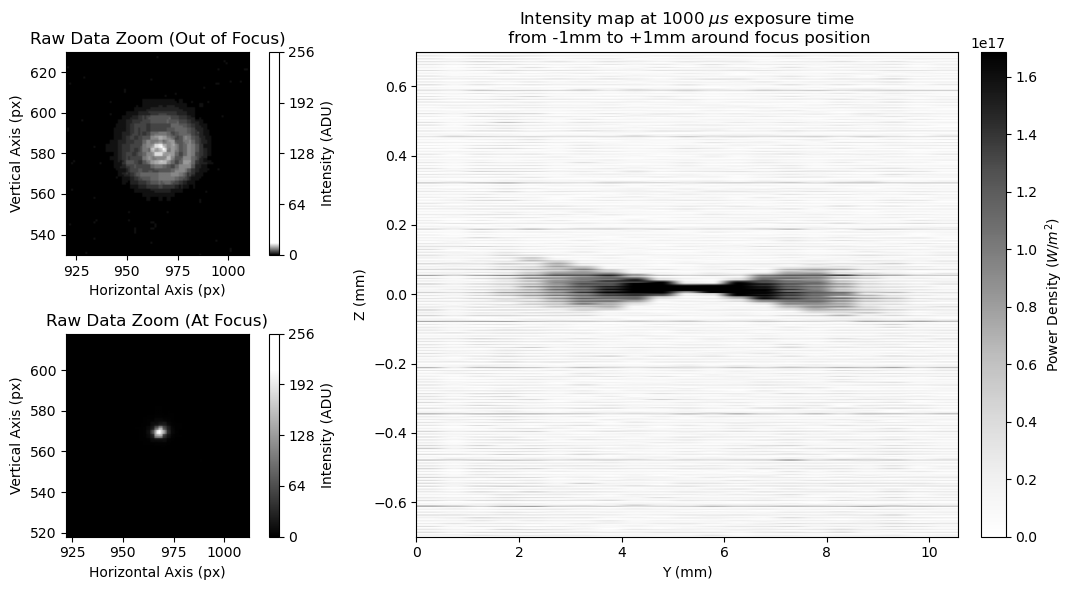}
        \caption{Experimentally reconstructed longitudinal intensity distribution ($X-Z$ cross-section) of the laser beam passing through the Fresnel zone plate focal volume.}
        \label{fig:focusscan}
    \end{minipage}
\end{figure}

\subsection{Theoretical Comparison and Focal Verification}

The reconstructed longitudinal profile (Figure \ref{fig:focusscan}) illustrates the formation of a highly symmetrical diffractive light cone on either side of the focal waist. This spatial evolution is compared directly with paraxial wave theory. 

Under the paraxial approximation, the convergence and subsequent divergence of the diffractive light cone are governed by the numerical aperture of the outermost zone \cite{goodman1996}. In our experimental prototype, the zone plate focal length $f_{exp}$ is identified by locating the plane of maximum axial intensity and optimal contrast, which corresponds to the narrowest transverse spot size in the through-focus profile \cite{miyakawa2014}:
\begin{equation}
    C_z = \frac{I_{\text{max}} - I_{\text{min}}}{I_{\text{max}} + I_{\text{min}}}
\end{equation}

By fitting this contrast curve to a Gaussian model, the plane of best focus $z^{*}$ is determined with high sub-pixel precision \cite{miyakawa2014}. 

The successful mapping of this focal waist demonstrates that our paraxial propagation framework—based on the discrete transfer function propagator \texttt{propTF}—accurately models not only simple refractive elements but also complex diffractive optics. This quantitative validation confirms that the numerical scaling conventions are fully preserved when translating physical experimental data into the computational domain. \cite{voelz2011}

\section{Iterative Phase Retrieval: The Gerchberg-Saxton Algorithm}

A fundamental limitation of physical optical sensors is that they are sensitive only to time-averaged irradiance (intensity) $I(x,y) \propto |E(x,y)|^2$, completely discarding the phase of the complex amplitude field. However, the phase distribution contains vital information regarding wavefront aberrations, propagation direction, and spatial coherence.\cite{miyakawa2014} While digital holography recombines a reference wave to record phase interferometrically, iterative phase retrieval represents a powerful, referenceless alternative that extracts the complex phase digitally from intensity measurements alone.\cite{zalevsky1996}

The Gerchberg-Saxton (GS) algorithm, originally proposed by Gerchberg and Saxton in 1972 for electron microscopy, is an iterative optimization framework that reconstructs a complex field by cycling between two planes with known intensity constraints. While classical GS algorithms propagate between the spatial domain and its Fourier transform plane, modern wave-optics metrology frequently employs the algorithm in the Fresnel (or diffractive) domain \cite{zalevsky1996}. Propagating between two axially separated diffractive planes (such as a focus and a defocused plane) significantly accelerates the convergence rate and reduces reconstruction error.\cite{zalevsky1996}

\subsection{Mathematical Formulation and Convergence Metrics}

We define two measurement planes along the optical propagation axis: Plane A (located at position $z_A$, representing the focal plane) and Plane B (located at $z_B$, representing a defocused plane), separated by an axial distance $\Delta z = z_B - z_A$. The known physical constraints are the measured intensities $I_A(x,y)$ and $I_B(x,y)$ captured at these planes.

The target amplitudes $A_A(x,y)$ and $A_B(x,y)$ are defined as the square roots of the corresponding irradiances \cite{zalevsky1996}:
\begin{equation}
    A_A(x,y) = \sqrt{I_A(x,y)}, \qquad A_B(x,y) = \sqrt{I_B(x,y)}
\end{equation}

If an initial phase estimate is unavailable, the algorithm begins by assigning a random phase distribution $\phi_{\text{init}}(x,y) \in [-\pi, \pi]$ to the amplitude at Plane A \cite{zalevsky1996}. Let $g_A^{(k)}(x,y)$ represent the estimated complex field at Plane A during iteration $k$. The iterative loop proceeds through the following sequential operations:

\begin{enumerate}
    \item \textbf{Enforce Plane A Constraints:} We construct the starting field estimate at Plane A by combining the known target amplitude $A_A$ with the retrieved phase $\phi_A^{(k)}$ \cite{zalevsky1996}:
    \begin{equation}
        g_A^{(k)}(x,y) = A_A(x,y) \exp\left(j \phi_A^{(k)}(x,y)\right)
    \end{equation}
    
    \item \textbf{Forward Fresnel Propagation:} The field is propagated forward by the distance $\Delta z$ to Plane B using the paraxial transfer function operator $\mathcal{P}$ \cite{zalevsky1996, voelz2011}:
    \begin{equation}
        G_B^{(k)}(x,y) = \mathcal{P}\left\{ g_A^{(k)}(x,y); \Delta z \right\} = \mathcal{F}^{-1}\left\{ \mathcal{F}\left\{g_A^{(k)}\right\} \cdot H(f_x, f_y; \Delta z) \right\}
    \end{equation}
    
    \item \textbf{Enforce Plane B Constraints:} The calculated amplitude is replaced by the target amplitude $A_B$, while preserving the phase $\angle G_B^{(k)}$ resulting from propagation \cite{zalevsky1996}:
    \begin{equation}
        g_B^{(k)}(x,y) = A_B(x,y) \exp\left(j \angle G_B^{(k)}(x,y)\right)
    \end{equation}
    
    \item \textbf{Backward Fresnel Propagation:} The modified field is propagated backward by the distance $-\Delta z$ to return to Plane A \cite{zalevsky1996}:
    \begin{equation}
        G_A^{(k+1)}(x,y) = \mathcal{P}\left\{ g_B^{(k)}(x,y); -\Delta z \right\} = \mathcal{F}^{-1}\left\{ \mathcal{F}\left\{g_B^{(k)}\right\} \cdot H(f_x, f_y; -\Delta z) \right\}
    \end{equation}
    The phase of this back-propagated field, $\phi_A^{(k+1)} = \angle G_A^{(k+1)}$, is extracted to initialize the next iteration \cite{zalevsky1996}.
\end{enumerate}

To evaluate convergence, we define the root-mean-square (RMS) error of the intensity profile at each step.\cite{zalevsky1996} In our numerical model, this error is calculated as the normalized Euclidean norm of the difference between the propagated field's intensity and the target Plane B measurement:
\begin{equation}
    \text{RMS}^{(k)} = \frac{\left\| \left|G_B^{(k)}(x,y)\right|^2 - I_B(x,y) \right\|_2}{\left\| I_B(x,y) \right\|_2}
\end{equation}

\subsection{Numerical Implementation and Code Structure}

Listing \ref{lst:gerchberg_saxton} details the core Python loop of our Fresnel-domain Gerchberg-Saxton implementation. By utilizing our paraxial propagator \texttt{propTF}, the code simulates the propagation steps using Fourier domain filtering, tracking error convergence over 50 iterations.

\begin{lstlisting}[language=Python, caption={Core Gerchberg-Saxton Iterative Loop}, label={lst:gerchberg_saxton}]
# Indices for Plane A (best focus) and Plane B (defocused plane)
idx_A = 20  # Plane A index in the z_scan array
idx_B = 19  # Plane B index in the z_scan array

# Known target intensities and propagation distance
I_z   = np.abs(E_z[:, idx_A])**2 # Intensity at Plane A (target)
I_z_p = np.abs(E_z[:, idx_B])**2 # Intensity at Plane B (target)
dz = z_scan[idx_B] - z_scan[idx_A] # Propagation distance (Delta z)

# Target amplitudes derived from intensities
amp_A = np.sqrt(I_z)
amp_B = np.sqrt(I_z_p)

# Initial phase guess at plane A: use Ef phase if available, otherwise random
try:
    init_phase = np.angle(Ef)
    if init_phase.size != amp_A.size:
        raise Exception()
except Exception:
    init_phase = np.random.uniform(-np.pi, np.pi, size=amp_A.size)

u_A = amp_A * np.exp(1j * init_phase)


num_iterations = 50
errors = np.zeros(num_iterations)

for k in range(num_iterations):
    # 1. Forward propagate from Plane A to Plane B
    u_B = propTF(u_A, Dx_m, wavelength_m, dz)
    
    # 2. Enforce amplitude constraint at Plane B
    u_B = amp_B * np.exp(1j * np.angle(u_B))
    
    # 3. Backward propagate from Plane B to Plane A
    u_A = propTF(u_B, Dx_m, wavelength_m, -dz)
    
    # 4. Enforce amplitude constraint at Plane A
    u_A = amp_A * np.exp(1j * np.angle(u_A))
    
    # 5. Monitor convergence: calculate normalized RMS error at Plane B
    recon_B = propTF(u_A, Dx_m, wavelength_m, dz)
    err = np.linalg.norm(np.abs(recon_B)**2 - I_z_p) / (np.linalg.norm(I_z_p) + 1e-20)
    errors[k] = err

# Final reconstructed fields at both planes
recon_A = u_A
recon_B = propTF(recon_A, Dx_m, wavelength_m, dz)
\end{lstlisting}

\subsection{Convergence Analysis and Experimental Considerations}

\begin{figure}[htbp]
    \centering
    \includegraphics[width=1\textwidth]{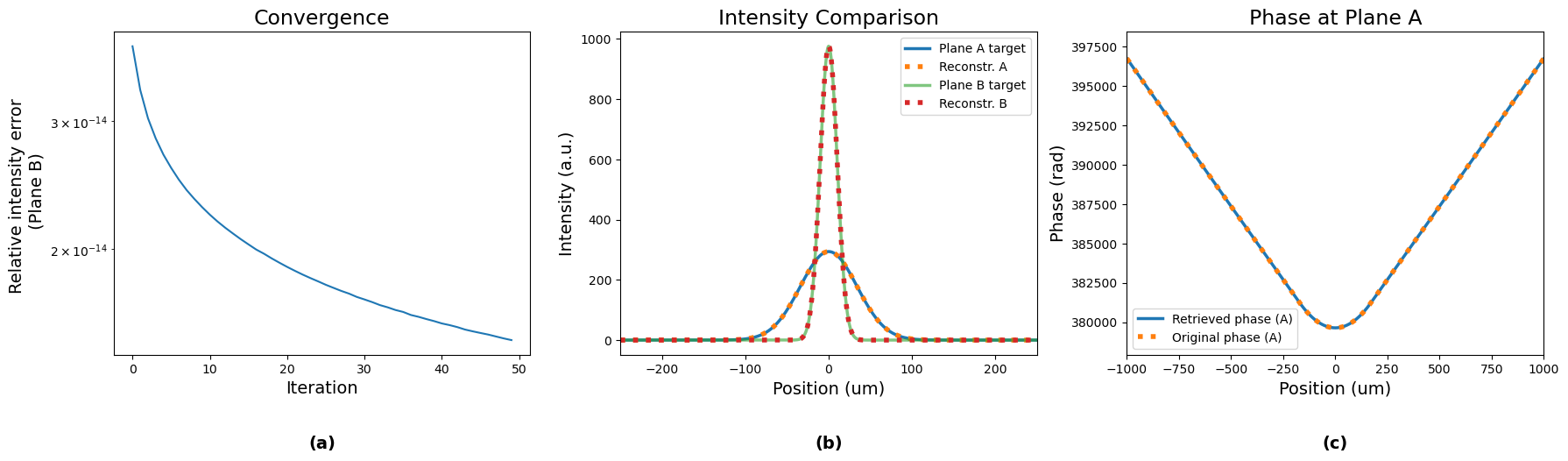}
    \caption{Fresnel-domain GS algorithm performance: (Left) Monotonic convergence of the relative intensity error over 50 iterations. (Center) Comparison of target and reconstructed intensities at both Plane A and Plane B. (Right) Retrieved phase versus the original phase at Plane A.}
    \label{fig:gs_results}
\end{figure}

The convergence performance of our numerical implementation is plotted in Figure~\ref{fig:gs_results}(a). In the absence of random noise, the relative intensity error decreases monotonically, reaching a machine-precision floor of approximately $10^{-14}$ within 50 iterations. The intensity overlay (Figure~\ref{fig:gs_results}(b)) demonstrates that the reconstructed fields converge perfectly to the target Gaussian profiles. Most importantly, the phase retrieval successfully recovers the original quadratic phase profile at Plane A, establishing a perfect overlay with the unwrapped phase of the physical field (Figure~\ref{fig:gs_results}(c)).

The diffractive GS framework provides an exceptionally stable and precise tool for wavefront characterization and beam diagnostics, matching analytical wave theory without requiring complex interferometric hardware \cite{zalevsky1996, miyakawa2014}.

\section{Structured Wavefronts: Helical-Wavefront Beams and Spiral Phaseplates}

While the diffractive elements discussed above focus optical energy into highly localized spots \cite{miyakawa2014}, wave optics also enables the generation of structured fields with custom intensity and phase topologies. A prominent class of structured light is the helical-wavefront laser beam—or spiral beam—which is characterized by a phase term proportional to the azimuthal angle $\theta$ \cite{zalevsky1996}:
\begin{equation}
    U(r, \theta, z) \propto \exp(j \ell \theta)
\end{equation}
where $\ell$ is an integer representing the topological charge of the vortex \cite{zalevsky1996}. At the center of the beam ($r = 0$), the phase has a screw dislocation (or phase singularity) where the phase is mathematically undefined \cite{zalevsky1996, goodman1996}. To satisfy the scalar wave equation, the complex field amplitude must vanish at this singularity, forming a dark central core surrounded by a bright ring—the classic "doughnut" intensity profile \cite{zalevsky1996, voelz2011}.

These fields represent orbital angular momentum (OAM) eigenmodes carrying a well-defined torque of $\ell\hbar$ per photon, completely independent of the wave's polarization state \cite{zalevsky1996}.

\subsection{Numerical Construction of the Vortex Phase}

Following discrete grid conventions, we represent this helical phase profile numerically by constructing a two-dimensional spatial Cartesian coordinate grid $(X,Y)$ using \texttt{np.meshgrid} \cite{voelz2011}. The azimuthal angle $\theta$ is computed at each pixel using the four-quadrant inverse tangent function (\texttt{np.arctan2}). The complex vortex field is then generated by multiplying a centered Gaussian amplitude distribution $u_{in}(x,y)$ by the spiral phase mask.

\begin{lstlisting}[language=Python, caption={Numerical Generation of a Charge-2 Vortex Beam}, label={lst:vortex_generation}]
# Generate a spiral (vortex) phase and propagate the vortex beam
charge = 2 # topological charge (ell)

# Compute azimuthal angle theta using a 2D meshgrid
X, Y = np.meshgrid(x_m, y_m)
theta = np.arctan2(Y, X)

# Create spiral phase mask and apply to the input Gaussian field
spiral_phase = np.exp(1j * charge * theta)
vortex = uin_V_m * spiral_phase
\end{lstlisting}

\begin{figure}[htbp]
    \centering
    \includegraphics[width=0.8\textwidth]{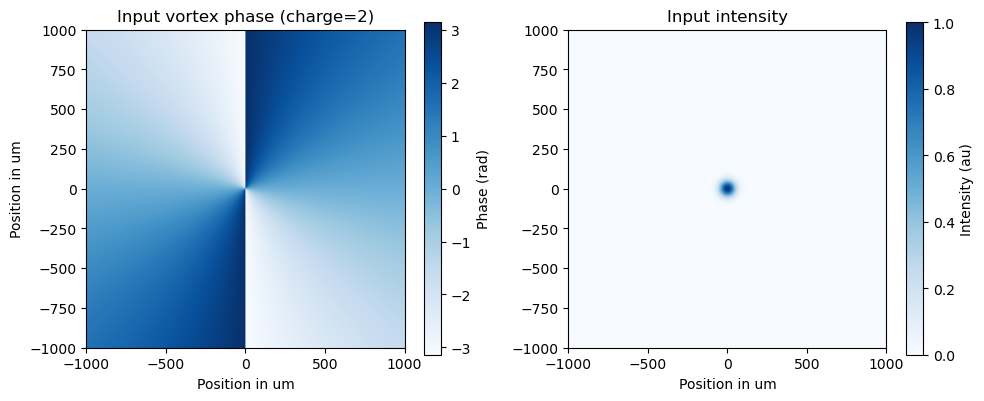} 
    \caption{Initial vortex beam state (charge l = 2): (Left) Azimuthal phase ramp showing 4π of total
phase rotation. (Right) Gaussian intensity profile immediately before propagation.}
    \label{fig:vortex_pt1}
\end{figure}

Immediately after the phase mask is applied ($z = 0$), the beam's intensity profile remains a solid, unaltered Gaussian spot because the phase modification does not affect the local irradiance $|U(x,y)|^2$  (Figure \ref{fig:vortex_pt1}) \cite{zalevsky1996, voelz2011}.

\subsection{Propagation and Laguerre-Gaussian Mode Decomposition}

The physical properties of the vortex beam emerge during free-space propagation. Using the paraxial transfer function propagator \texttt{propTF}, the field is evolved over a distance $z = 0.05\text{ m}$.

\begin{lstlisting}[language=Python, caption={Propagation of the Vortex Field}, label={lst:vortex_propagation}]
# Define propagation parameters
z_prop_m = 0.05
wavelength = 500e-9

# Compute the complex field at distance z
field_out = propTF(vortex, Dx_m, wavelength, z_prop_m)
\end{lstlisting}

Because a ``spiralized" Gaussian beam is no longer a pure transverse mode of free space, it is best analyzed by decomposing the exiting field into a complete set of orthogonal Laguerre-Gaussian (LG) modes \cite{zalevsky1996}. LG modes propagate with only a scale change of their intensity distribution, and they naturally accommodate helical phase terms of the form $\exp(j \ell \theta)$ \cite{zalevsky1996}.

\begin{figure}[htbp]
    \centering
    \includegraphics[width=0.8\textwidth]{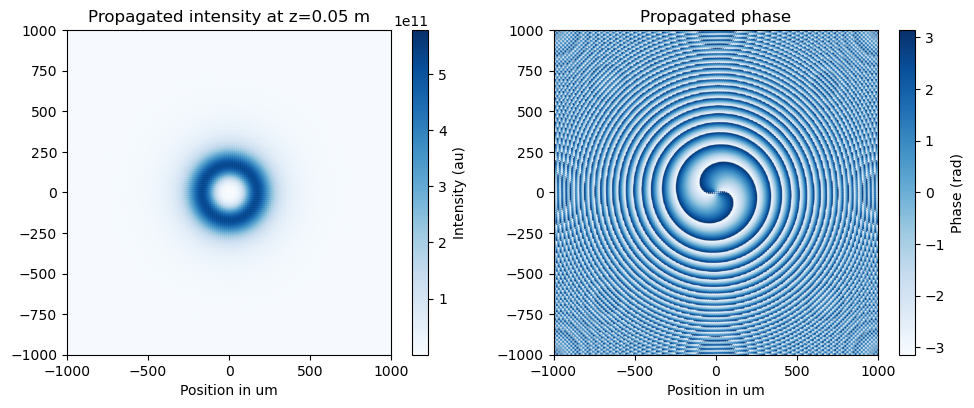} 
    \caption{Propagated vortex beam at z = 0.05 m: (Left) Intensity distribution showing the central
singularity. (Right) Helical phase structure showing the spiral evolution of the wavefront.}
    \label{fig:vortex_pt2}
\end{figure}

When propagating downstream, the destructive interference of these superimposed modes close to the optical axis enforces the on-axis intensity null \cite{zalevsky1996}. This results in the ring-like "doughnut" intensity profile and helical spiral phase arms illustrated in Figure \ref{fig:vortex_pt2}.

\begin{figure}[htbp]
    \centering
    \includegraphics[width=0.8\textwidth]{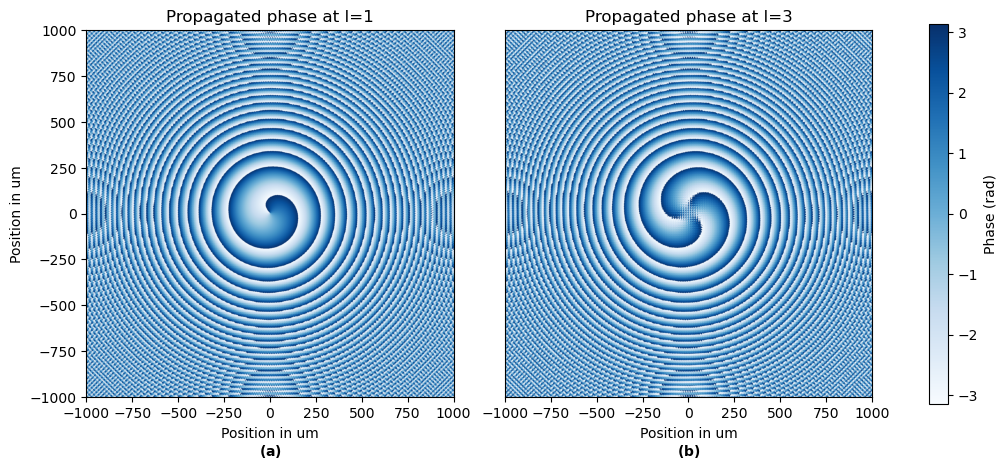} 
    \caption{(a) Just as the $l=2$ phase plot had two ``arms," at $l=1$ the propagated phase has only one "arm." (b) At $l=3$ the propagated phase has three.}
    \label{fig:vortex_pt3}
\end{figure}

By varying the \texttt{charge} parameter, the framework can simulate a variety of structured light fields, vortices of ($l=1$) (Figure \ref{fig:vortex_pt3}(a)) or ($l=3$) (Figure \ref{fig:vortex_pt3}(b)). 

\section{Conclusion}
In this work, we presented a comprehensive computational and experimental framework for wave optics and scalar diffraction modeling implemented entirely in Python. Heavily inspired by David Voelz's foundational text, \textit{Computational Fourier Optics: A MATLAB Tutorial} \cite{voelz2011}, this project translates classical numerical Fourier optics algorithms—originally developed for the MATLAB environment—into an open-source, flexible, and accessible Python framework.

Throughout this paper, we established the foundational numerical building blocks required for discrete wave-optics simulations, beginning with 2-D spatial grid conventions, frequency-domain array centering, and paraxial transfer-function propagation (\texttt{propTF}). We demonstrated the modeling of both refractive optical systems—such as positive thin lenses and conjugate magnification scaling—and diffractive optical elements, including binary amplitude gratings, Fresnel zone plates, and structured optical vortex beams carrying orbital angular momentum. Furthermore, we validated these theoretical tools against laboratory through-focus profiling data and demonstrated advanced wavefront recovery methods, including phase-only digital holography playback and referenceless, iterative Fresnel-domain phase retrieval using the Gerchberg-Saxton algorithm.

To streamline future work and make these tools readily accessible to the broader optics community, the core functions developed throughout this research—including spatial grid generators, transfer-function propagators, diffractive mask constructors, and phase-retrieval routines—have been bundled into a dedicated open-source Python package. The \texttt{topolino} package is published on the Python Package Index (PyPI) and can be accessed at \url{https://pypi.org/project/topolino/}. By bridging classical scalar wave theory with modern Python scientific computing libraries, this work provides a versatile foundation for optical design, beam diagnostics, and computational metrology.

\newpage

\bibliography{fourier.bib} 
\bibliographystyle{iopart-num} 

\end{document}